\documentclass[journal,comsoc]{IEEEtran}
\pdfoutput=1 
\usepackage{array}
\usepackage[T1]{fontenc}
\usepackage{pifont}
\usepackage{comment}
\usepackage{soul,color}
\usepackage{tikz}
\usepackage{url}
\usepackage{tabularx}
\usepackage{multirow}
\usepackage{color}
\usepackage{caption}
\usepackage{xcolor}
\usepackage{hhline}
\usepackage{colortbl}
\usepackage{cite}
\usepackage{amsmath,amssymb,amsfonts}
\usepackage{algorithmic}
\usepackage{graphicx}
\usepackage{textcomp}
\usepackage{subfig}
\usepackage{float}
\usepackage{adjustbox}

\graphicspath{ {./figures/} }

\usepackage{url}
\begin{document}

\title{A Comprehensive Study of Virtual Machine and Container Based Core Network Components Migration in OpenROADM SDN-Enabled Network}

\author{Shunmugapriya~Ramanathan, Koteswararao~Kondepu, Tianliang~Zhang,
        Behzad~Mirkhanzadeh, \\
        Miguel~Razo,
        Marco~Tacca,
        Luca~Valcarenghi,
         and~Andrea~Fumagalli
\thanks{S. Ramanathan, T. Zhang, B. Mirkhanzadeh, M. Razo, M. Tacca, and A. Fumagalli are with Erik Jonsson School of Engineering and Computer Science, Open Networking Advanced Research (OpNeAR) Lab, The University of Texas at Dallas, Richardson, TX, USA.}
\thanks{K. Kondepu is with Indian Institute of Technology Dharwad, Dharwad, India.}
\thanks{L. Valcarenghi is with Scuola Superiore Sant'Anna, Pisa, Italy.}
\thanks{The updated version published at IEEE Access.}}

\clearpage
\maketitle
\thispagestyle{empty}
\pagestyle{empty}
\vspace{-1.2em}
\begin{abstract}
With the increasing demand for openness, flexibility, and monetization the Network Function Virtualization (NFV) of mobile network functions has become the embracing factor for most mobile network operators. Early reported field deployments of virtualized Evolved Packet Core (EPC) --- the core network component of 4G LTE and 5G non-standalone mobile networks --- reflect this growing trend. To best meet the requirements of power management, load balancing, and fault tolerance in the cloud environment, the need for live migration for these virtualized components cannot be shunned. Virtualization platforms of interest include both Virtual Machines (VMs) and Containers, with the latter option offering more lightweight characteristics.

The first contribution of this paper is the implementation of a number of custom functions that enable migration of Containers supporting virtualized EPC components. The current CRIU-based migration of Docker Container does not fully support the mobile network protocol stack. CRIU extensions to support the mobile network protocol stack are therefore required and described in the paper. The second contribution is an experimental-based comprehensive analysis of live migration in two backhaul network settings and two virtualization technologies. The two backhaul network settings are the one provided by CloudLab and one based on a programmable optical network testbed that makes use of OpenROADM dense wavelength division multiplexing (DWDM) equipment. The paper compares the migration performance of the proposed implementation of OpenAirInterface (OAI) based containerized EPC components with the one utilizing VMs, running in OpenStack.

The presented experimental comparison accounts for a number of system parameters and configurations, image size of the virtualized EPC components, network characteristics, and signal propagation time across the OpenROADM backhaul network. The comparison reveals that the live migration completion time of the virtualized EPC components is shorter in the Container platform, while the service downtime is shorter in the VM OpenStack platform. Fine tuning of key parameters may be required for optimal performance.

\end{abstract}

\vspace{-0.1cm}
\section{Introduction}
\vspace{-0.1cm}
\IEEEPARstart{T}{he} 3GPP standards for the 5G mobile communication 
and the ESTI NFV~\cite{NFV} are two key enablers for 5G virtualization. The Non-Standalone version of the 5G mobile communication comprises the New Radio (NR) and the Next Generation Radio Access Network (NG-RAN), including the gNodeB (gNB), connected to the 4G EPC. In the NG-RAN system, the gNB is disaggregated into three components, namely the Remote Radio Unit (RRU), the Distributed Unit (DU), and the Central Unit (CU)~\cite{IMT2020}. The RRU mainly comprises the RF components, while the functions performed by the DU and the CU vary based on the chosen split option from the available list in the 3GPP standards~\cite{f_splits}. 

The use of both commercial off-the-shelf (COTS) hardware and Network Function Virtualization (NFV) helps Mobile Network Operators (MNO) reduce their operational cost and the need for excessive over-provisioning of network capacity in order to achieve the much needed support for service redundancy~\cite{NFV-Part1,NFVSDN}. Well-established virtualization platforms exist that support both the Virtual Machine (VM) and Container based hardware virtualization such as OpenStack, and VMware~\cite{OpenStack, VMwar}.
VMs can concurrently and independently run on the same host compute hardware while each provides a distinct OS support to its guest application, namely each VNF. Docker makes use of OS-level virtualization to produce VNFs that run in packages called Containers. Container-based solutions have been gaining traction in the recent years due to their reduced overhead. While these platforms are widely used, some open challenges still remain to be addressed~\cite{VirtChallenge}. One of these challenges is to achieve the required carrier-grade Service Level Agreement (SLA) in the virtualization platform that supports NFV~\cite{Wang2014CarrierGradeDC}. It is believed that the NFV compute platform must utilize the fullest features such as (live) migration, snapshot, and rebirth in order to ensure that the SLA requirements are finally met in terms of security, reliability, and total cost of ownership. 

Focusing on the first feature, live migration is the process of migrating the VNFs from one host to another while guaranteeing zero or minimal impact to the connectivity service offered to the mobile network users. Being able to live migrate VNFs offers a number of significant advantages. VNFs can be moved away from overloaded servers/hosts and reallocated in less loaded compute nodes. Load balancing in the compute nodes~\cite{LoadBalance} can be timely achieved by redistributing VNFs to sparsely loaded servers. 
To effectively perform maintenance --- such as upgrading OS versions and changing network configurations --- or fault management, live migration of VNFs is often required. Last but not least, cost savings in terms of power consumption management may be facilitated through VNF migration. For example, when some servers are underutilized for a prolonged period of time, their VNFs may be relocated elsewhere to allow these computing elements to be shutdown.

Realizing the importance of timely offering virtualized EPC solutions with built-in capability for VNF live migration, this paper describes a few experimental settings designed to achieve this goal.
These experimental settings are obtained by leveraging open software and standard 
solutions whenever possible, and by implementing additional custom software 
packages when that is necessary to complete the required NFV/SDN architecture. All hardware components are commercially available. The ultimate objective is to validate the feasibility and compare the performance of a few plausible NFV/SDN architectures, which provide live migration of EPC virtualized functions with minimal connectivity disruption to the mobile user. Specifically, the VNFs for which live migration is tested are three core network components, namely Home Subscriber Server (HSS), Mobility Management Entity (MME), and Serving and Packet Gateway (SPGW). These virtualized EPC components are implemented using the OpenAirInterface (OAI) software package.

Two virtualization technologies are considered, one based on VMs and the other based on Docker Containers. In the former platform, live migration of VNFs running as VMs is achieved through Kernel-based Virtual Machine$\slash$Quick EMUlator (KVM$\slash$QEMU) with the \emph{libvirt} API~\cite{Hypervisor}. In the latter platform, live migration of VNFs running as Docker Containers is achieved through Checkpoint and Restore In Userspace (CRIU)~\cite{CRIU}.

It must be noted that the currently available CRIU software package
does not offer two key functionalities~\cite{CRIULimitation} that are required to support Container-based VNF migration in the C-RAN backhaul network, which are:
i) support for the Stream Controlled Transmission Protocol (SCTP), 
used in the LTE network to guarantee message delivery between MME and CU; and
ii) GTP (GPRS Tunnelling Protocol) device-specific information needed by the SPGW software to provide tunnelling of the user data traffic. 
To overcome these limitations of the CRIU, two custom solutions
described in Section~\ref{sec:dc}, have been implemented and integrated in the experimental settings, which are: i) support for SCTP in the CRIU software and ii) a utility software to handle the GTP device-specific information.

The two virtualization technologies are tested in two distinct experimental settings. In the first setting, compute hosts are realized using repurposed Stampede~\cite{TACC} servers to form two geographically distinct edge compute sites.
The two sites are connected through a backhaul fiber-optics network that is realized using reconfigurable optical add-drop multiplexing (ROADM) equipment,
optical transponders, and switchponders from a number of equipment manufacturers. The optical equipment is OpenROADM compliant~\cite{OpenROADM1} and controlled by the open source TransportPCE controller~\cite{TransPCE}.
Orchestration of resource allocation in the optical network, Ethernet switches, and Stampede compute nodes is provided by a custom PROnet Orchestrator software~\cite{OpenROADM1}. Both virtualization technologies are also tested using the CloudLab federated testbed~\cite{Duplyakin+:ATC19}, which provides an additional benchmark to validate the newly added custom code in an open environment. This second setting offers a more performing and diverse compute hardware platform compared to the first one but does not provide an optical backhaul network that can be controlled by the experimenter. 

\section{Related Work}

In~\cite{3gpp23}, the 3rd Generation Partnership Project (3GPP) specifies different resiliency mechanisms for EPC components, and handling failures with the help of Echo Request/Response timer messages. In addition,~\cite{Carpio} presents approaches for recovering VNF through replication and migration of network functions when outages affect compute resources. Moreover, infrastructure network failures can be recovered directly at the network level, for example by resorting to a Software Defined Network (SDN) controller~\cite{Giorgetti}, or by combining replication/migration with connection rerouting. In~\cite{JOCN2018}, a two-step resiliency scheme is proposed for RAN functional split reconfiguration by orchestrating lightpath transmission adaptation. 

Ref.~\cite{5GPP5GV} refers to the advantage of VNFs by conducting a survey and collecting technical inputs from the 5G-PPP projects. Most of the project prototypes evolved from ETSI MANO resort to OpenStack Virtualized Infrastructure Manager (VIM) with the addition of Kubernetes orchestration to host both containerized network functions and the classical VM-based VNFs. 

In~\cite{GopalasinghamIM2017}, the authors evaluate the  performance of the Virtualized RAN using both VM and Docker Container in the SDN enabled network. Using their analytical model and experimentation, they report that Docker Container performance is superior compared to VMs in terms of IO performance and communication latency. The authors analyse the service rate, average waiting time, inter-arrival time for both the VMs and Docker Container using a queueing model. However, the aspects concerning migration techniques and related implementation challenges in different virtualization technologies are not addressed in~\cite{GopalasinghamIM2017}.

The VNF migration of virtualized CU$\slash$virtualized DU (vCU$\slash$vDU) over WDM network using CRIU is briefly discussed in~\cite{FengOFC2020}. Here, the authors mention checkpointing the vCU by collecting the CPU state and memory page information and storing them on disk. The collected metadata is restored at the destination host by the lightpath reconfiguration to ensure the connectivity of the end-user. 

So far, there are no papers that address implementation and provide detailed evaluation of NFV-SDN systems performing live migration of VM and Container supporting core network functions. 

\section{Live Migration of Virtualized Core Network functions: Techniques, Limitations, and Solutions}

This section describes the VNF migration strategies exploited, the key implementation challenges faced, and the custom software solutions developed for successfully performing core network components live migration.

\subsection{Docker Container Migration with CRIU} 
\label{sec:dc}

With the Container technology gaining increasing attention due to its smaller footprints, many recent research efforts focus on designing the optimal Container migration approach, where the VNF memory, file system, and network connectivity state need to be transferred to the destination host without disrupting the running application~\cite{ContPerf,ContPerf1}. A number of prominent Container run-time software packages handles the migration in user space through checkpoint and restoration technique.

\begin{figure}[htbp]
     \vspace{-1.1em}
     \centering
     \includegraphics[trim=0 0 0 -1cm,width=1.0\columnwidth]{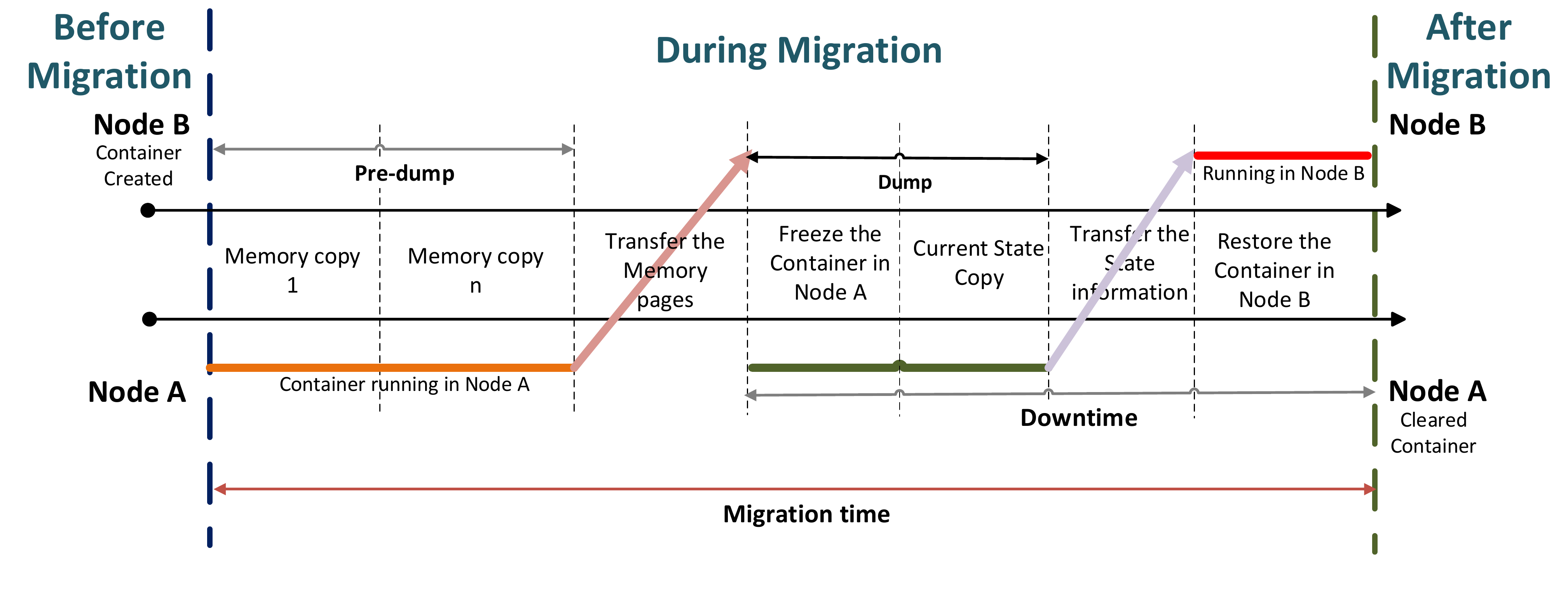}
    \caption{CRIU-based Container live migration}
    \label{fig:CRIU_Mig}
    \vspace{-1.2em}
    \vspace{1em}
\end{figure}

As shown in Fig.~\ref{fig:CRIU_Mig}, during checkpoint, the CRIU method freezes the running Container at the source node (host A) and collects metadata about the CPU state, memory content, and information about the process tree~\cite{pstree} associated with the running Container service.
The collected metadata information is passed on to the destination node (host B), and the restore process resumes the Container service from the frozen point with the copied metadata at the destination node. The total time required to perform the checkpoint, metadata copy and transmission, and Container restoration contributes to the frozen time of the application. There are ways to reduce the frozen time using lazy migration method~\cite{CRIULazy}. This method is however outside the scope of this study.

The following subsections describe the CRIU current limitations that are encountered during the specific migration of certain core network components~\cite{CRIULimitation} and how these limitations are circumvented in order to successfully perform such components' live migration.


\subsubsection{HSS Live Migration with CRIU Support}

A TCP connection is required between HSS and MME.
The HSS component establishes the TCP socket at start time and stores relevant user information in a MySQL database.
Upon performing migration of the HSS component, the CRIU application 
needs to copy the database information into the memory page and 
restore the TCP connectivity at the destination host without disturbing 
the peer end connection state at the MME side.
The \emph{tcp-established mode}~\cite{CRIUTCP} must be set in the CRIU configuration
in order to ensure TCP connection re-establishment at the destination host. 
This TCP repair mode is supported starting from version 3.5 of Linux Kernel mainline,
which provides support for socket re-establishment without requiring the exchange
of the initial TCP connection setup messages.
No additional custom software is required in CRIU. 

\subsubsection{MME Live Migration with CRIU Support}

The MME component makes use of SCTP (stream control transmission protocol) to exchange S1-MME messages with multiple gNBs.
One of the main differences between TCP and SCTP is that TCP has a single association in the given socket connectivity whereas SCTP has multiple associations with the single socket connection using the stream options.
The SCTP protocol is not supported in the currently available CRIU release version, 
and consequently migration of the MME component cannot be
executed successfully unless a new SCTP connection is re-established with a new start message handshake. 
To overcome this CRIU limitation it was necessary to design and develop the additional custom software described next. 
When the SCTP socket is in \emph{listen mode}, adding support in CRIU for SCTP is relatively simple because only user-space software changes are required.
When the endpoint association is in \emph{established mode}, the associativity endpoint details along with the kernel code changes are also needed, thus adding complexity to the required procedure. The authors developed a procedure in CRIU to support migration of one-to-one style of SCTP socket along with the required kernel changes for achieving automatic SCTP socket re-establishment at the destination host. The kernel code was modified in such a way that when the MME metadata information is passed onto the destination host, the kernel is able to re-establish the SCTP socket at the destination host without requiring to re-instantiate the SCTP connection. This feature is now available when the SCTP repair mode is turned on. 

\subsubsection{SPGW Live Migration with CRIU Support}
\label{sec:spgw_criu}

The SPGW component makes use of a GTP interface for handling the User Equipment 
(UE) connectivity and maintains the GTP tunnel list up to date with UEs and base station (gNB) relevant information. 
With the currently available open software platforms (OAI, CRIU) 
these critical pieces of information are not carried over onto the destination host during the SPGW migration.
Consequently the end-user connectivity is lost and the entire UE connection 
\emph{re-establishment} has to take place again starting from the base station.
For the reader's convenience a short overview of the GTP tunnel mechanism is first provided, 
followed by the description of the custom software that was developed to overcome this severe limitation.

\begin{figure}[htbp]
     \vspace{-1.1em}
     \centering
     \includegraphics[trim=0 0 0 -1cm,width=1.0\columnwidth]{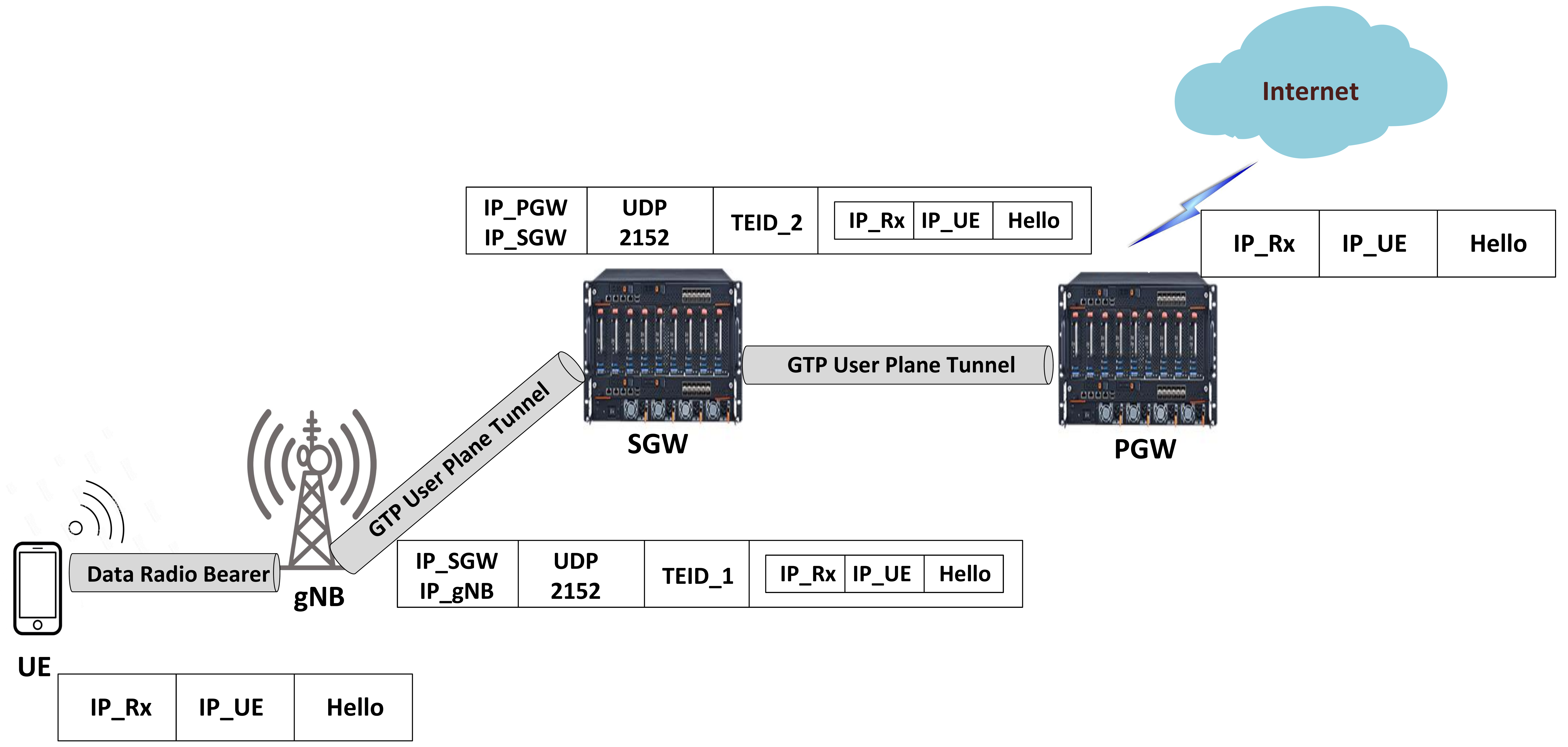}
    \caption{GTP uplink IP traffic from the UE to the Internet  }
    \label{fig:GTP_UDP_IP}
    \vspace{-1.2em}
    \vspace{1em}
\end{figure}

To provide mobility to the UE and cope with the resulting network topology dependencies, the UE uplink and downlink IP packets are routed through a GTP tunnel that is previously established between the base station and the SPGW. Tunnel Endpoint Identifier (TEID) values are mutually exchanged between the base station and SPGW to ensure correct flow of data traffic. 
For example, considering the UE uplink communication, the IP data packet is first encapsulated at the base station by adding its IP/UDP/GTP header and transmitted in the GTP tunnel to reach the Service Gateway (SGW). 
The SGW replaces the outer header with its IP/UDP/GTP header and sends it to the Packet Gateway (PGW). 
The PGW decapsulates the outer header, and passes the original UE IP data packet to the Internet/Packet Data Network (PDN). 
In this solution the base station acts as the Serving GPRS Support Node (SGSN) 
and the SPGW acts as the Gateway GPRS Support Node (GGSN). 
The GTP-U communication along with the UDP/GTP header addition is illustrated in Fig.~\ref{fig:GTP_UDP_IP}. 
The OAI SPGW software implements the above mentioned data plane connectivity by using the Linux Kernel GTP tunnelling module. This kernel module creates the GTP device interface (\emph{gtp0}) for tunnelling the user data traffic to the PDN, and the SPGW software maintains the SGSN TEID information in the GPRS tunnelling list. 

A procedure had to be identified to overcome the current CRIU software inability to store device interface-specific information. 
Additionally, with the CRIU software operating in user space, the kernel dependent \emph{gtp0} 
device interface information cannot be checkpointed. 
Finally, the tunnelling list associated with that interface (\emph{gtp0}) is completely lost too when the SPGW metadata information is moved to the destination host. In essence, the restored SPGW application at the destination host holds the UDP socket connectivity with the gNB endpoint, but without the required \emph{gtp0} interface and tunnel list, unfortunately. 
The authors had to upgrade the OAI software package with a software utility program that overcomes these current limitations.

The software utility program is implemented to run along with the SPGW container and the utility program takes care of 
i) creating the \emph{gtp0} interface thereby managing the rtnetlink socket~\cite{GGSN}; ii) reading and configuring the \emph{gtp0} interface-specific information from the SPGW configuration file (e.g., IP address, MTU size, mask, and routing information); 
iii) enabling masquerading for SGi interface --- point of interconnection between the PGW and the external network (PDN) --- by adding POSTROUTING iptable commands --- helps to alter the IP packets after routing completed; and iv) maintaining the GTP tunnel list information of the running SPGW Container application after migration. 
Thus, when the SPGW Container is checkpointed, along with the CRIU collected metadata, the utility program adds its \emph{gtp0} interface-specific information and the tunnel list data. During restore time, the SPGW application is restored successfully at the host with both the socket connectivity and the GTP related information to reinstate the end-user communication. Additional details about the OAI design changes that are applied to handle SPGW CRIU migration can be found in~\cite{PriICTON2020}.

\subsection{VM Migration with the KVM/QEMU Hypervisor}

A hypervisor is a software-based virtualization layer between the physical machine (host) and the VM guests running on it. The hypervisor takes care of scheduling and allocating compute resources to the VM guests. KVM hypervisor is a kernel module integrated with version 2.6.20 of mainline Linux Kernel that is used in OpenStack~\cite{Hypervisor} for providing the virtualization infrastructure. The QEMU-KVM module provides the VM management such as spawning and migrating VMs using the guest execution mode. Interaction with the KVM/QEMU hypervisor is made possible through the \emph{libvirt} library and its set of API calls. 

\begin{figure}[htbp]
     \vspace{-1.1em}
     \centering
     \includegraphics[trim=0 0 0 -1cm,width=1.0\columnwidth]{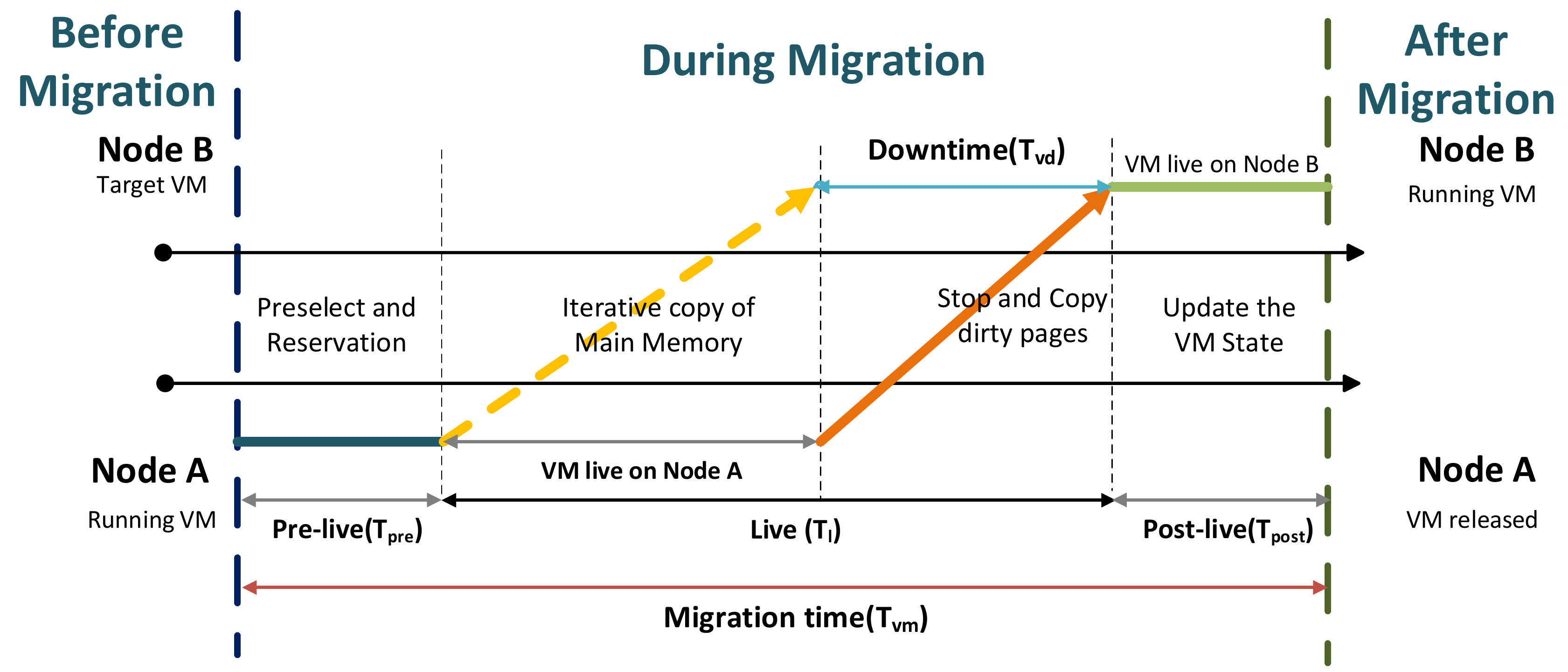}
    \caption{Pre-Copy based VM Live Migration.  }
    \label{fig:VM_Mig}
    \vspace{-1.2em}
    \vspace{1em}
\end{figure}

During VM migration, the CPU state, memory state, network and disk image of the entire VM are migrated from the source to destination host. During the memory pages coping process, the dirty pages (i.e., modified memory pages) are iteratively transferred -- referred to as \emph{push} phase, while the VM is still running at the source host. Once the maximum iteration count is reached, the VM is temporarily stopped at the source host, all the main memory pages are copied to the destination and then the VM is resumed at the destination host. This process of memory page coping --- referred to as \emph{Pre-copy} method --- is illustrated in Fig.~\ref{fig:VM_Mig}. 

There exists another coping strategy named \emph{Post-copy}~\cite{Pre_PostCopy}, where the stop and copy phase happens first so that the VM is started earlier at the destination host. Then the remaining dirty pages are copied at the time of page fault occurrence, a technique that is referred to as \emph{pull} phase. Only the \emph{Pre-copy} migration method is used in the experiments discussed in this paper, since it is an optimized method for memory read intensive applications. 
As shown in Fig.~\ref{fig:VM_Mig}, the \emph{Pre-live} phase considers the preselect and reservation process such as preparing the destination host with the VNF instance details of keypair association and network information, the \emph{live} phase carries the memory page copy process to the destination host -- pull phase for the Pre-Copy method and the \emph{Post-live} phase performs post operation after the live migration such as it updates the running VM state in the MySQL database and Neutron database with the host information and port details. 

\subsubsection{VM Migration Limitations Handling} 
Due to the KVM$\slash$QEMU hypervisor maturity as virtualization technology~\cite{VMStudy}, the VM components running core network elements can be more easily migrated when compared to Docker Containers.
Only a few precautions are necessary to ensure correct migration execution in the OpenStack environment,
where layer-2 network connectivity is provided by default using the Open Virtual Switch (OVS) integration bridge~\cite{OVS}. 
Even though the OpenStack security rule permits the use of the SCTP protocol, the OVS firewall blocks the SCTP packets preventing them from reaching the hosts. To circumvent this drawback, SCTP messages are encapsulated inside UDP frames using Open Virtual Private Network$\slash$Virtual Extensible LAN (OpenVPN$\slash$VXLAN) connectivity. The resulting UDP frames are therefore not blocked by the OVS firewall. In summary, the SCTP protocol communication between MME and CU is made possible through an OpenVPN service enabled in the S1-MME communication interface~\cite{PriICTON2019}.

\section{Experiment Testbeds}
\label{sec:et}
Two testbeds are used in this study, which provide all of the required
C-RAN system key components including radio hardware units, compute nodes, Ethernet switches, and optical transport network equipment.
The first testbed is implemented at the University of Texas at Dallas (UTD)
and makes use of optical transport network equipment that is OpenROADM compliant~\cite{OpenROADM}.
The second testbed consists of ClouldLab~\cite{Duplyakin+:ATC19} compute resources connected to the radio units hosted at UTD through the Internet. Combined, the two testbeds provide an opportunity to test the described procedures to live migrate virtualized core network components in the presence of state-of-the-art programmable optical network equipment on the one hand (in the former testbed) 
while also ensuring compliance of the proposed software implementation in an open federated environment on the other (in the latter testbed).

In both testbeds, the C-RAN software modules are implemented using OAI~\cite{OAI2014}, while the radio hardware units are implemented using NI B210 radio prototyping boards~\cite{USRP}, as shown in Figs.~\ref{fig:OpenROADM_Testbed} and \ref{fig:Cloudlab_Testbed}. The OAI software version considered for the CU and DU is v2019.w25 and for the core network, v0.5.0-4-g724542d is used. The radio hardware unit and DU interface is realized using the USB 3.0 (B210 radio) interface with the DU running on a dedicated physical machine. All of the experiments make use of option 2 split~\cite{3GPPTS} between DU and CU, according to which both Packet Data Convergence Protocol (PDCP) and Radio Resource Control (RRC) services run on the CU.
 
\subsection{Testbed 1: OpenROADM}
\label{sec:oroadm}

\begin{figure*}[ht]
\centering
\includegraphics[width=2.0\columnwidth]{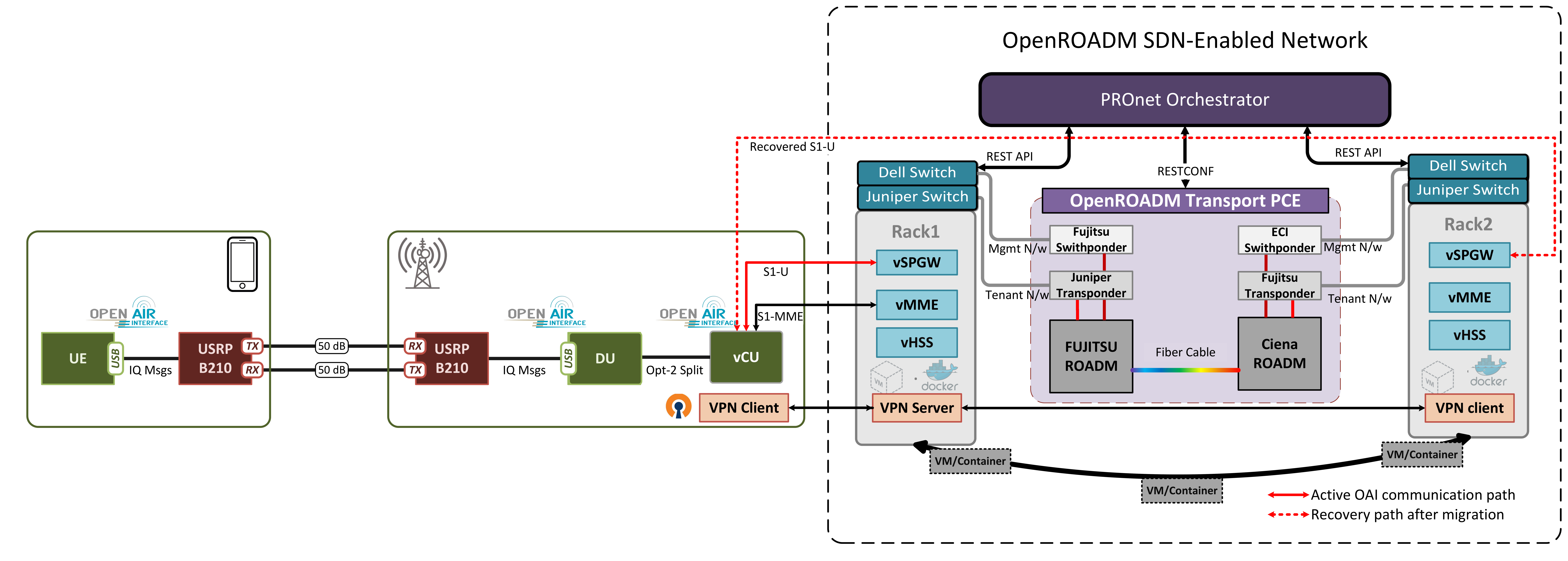}
\caption{The OpenROADM testbed~\cite{ONDM2021}.}
\label{fig:OpenROADM_Testbed}
\end{figure*}

Fig.~\ref{fig:OpenROADM_Testbed} shows the block diagram of the OpenROADM testbed configuration 
used to investigate both the KVM and CRIU based migration procedures. 
Two racks of Stampede compute nodes are connected through
an optical transport (backhaul) network comprising only OpenROADM compliant equipment.
The virtualized EPC software components (HSS, MME, SPGW) are first executed on the left rack (Rack 1).
Once triggered, the live migration of either the VM or Container that supports one of these EPC components 
takes place over a dedicated optical circuit (lightpath) that is dynamically created between the two racks 
to form a temporary high-speed connection in the management network to expedite the migration procedure between racks.
The optical transport (backhaul) network consists of two OpenROADM nodes provided 
by Ciena (6500) and Fujitsu (1FINITY) for routing lightpaths between the two racks or compute sites.
Transmission and reception of Ethernet client signals across the optical transport network 
are realized by deploying OpenROADM compliant
Fujitsu (1FINITY) T300 100G Transponder and Juniper ACX6160-SF Transponder for the tenant network, 
and Fujitsu (1FINITY) F200 1G/10G/100G Switchponder and ECI Apollo OTN OpenROADM switchponder 
for the management network.
The optical equipment is controlled by the open source optical network controller TransportPCE version 2.0.0, which is an application running on OpenDaylight version 6.0.9.
Also shown in Fig.~\ref{fig:OpenROADM_Testbed}, the programmable optical network (PROnet) Orchestrator is a software module 
developed at UTD to coordinate automatic resource provisioning in an Ethernet-over-WDM network~\cite{PROnetBM2018}. OpenFlow~\cite{OpenFlow, OpenNetwork} enabled switches (Juniper QFX5120 and Dell N3048p) --- controlled by the PROnet Orchestrator --- are used to interconnect compute nodes in the two racks and also to route packets (in both management and tenant networks) to the assigned transport optical equipment.
The PROnet Orchestrator was recently upgraded with two additional features~\cite{OFCBM2020}: a RESTCONF interface to work with the TransportPCE northbound API which relies on the OpenROADM Service Model,
and a REST API to work with OpenStack. With these two upgrades, the PROnet Orchestrator offers a single point of control and coordination of the compute and network resources in the described experimental setting.
For example, to enable experimentation with varying backhaul network round trip delays, the PROnet orchestrator is instructed to create lightpaths in the OpenROADM network with varying end-to-end propagation distances, i.e., a few meters -- considered as short distance, 25~km, and 50~km.
During the migration process, the PROnet Orchestrator first triggers the creation of the management lightpath between the two racks and then initiates the migration of one of the EPC virtual components. The migration procedure is carried out through the OpenStack dashboard when using VMs and through shell script commands when using Containers. 
\subsection{Testbed 2: CloudLab}
\begin{figure}[htbp]
     \vspace{-1.1em}
     \centering
     \includegraphics[trim=0 0 0 -1cm,width=1.0\columnwidth]{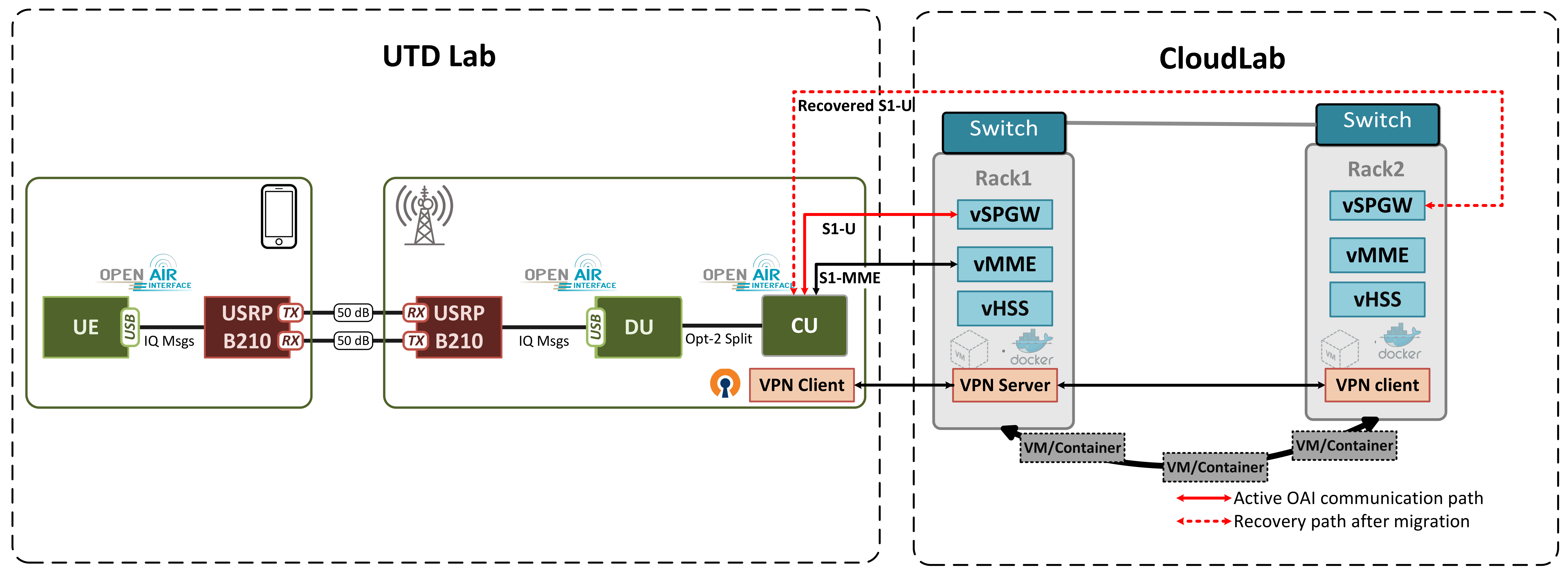}
    \caption{The CloudLab testbed~\cite{Access2021}.}
    \label{fig:Cloudlab_Testbed}
    \vspace{-1.2em}
\end{figure}

Fig.~\ref{fig:Cloudlab_Testbed} shows the block diagram of the CloudLab testbed,
in which some of the RAN components --- UE, DU, and CU --- reside in the UTD Lab while the virtualized core network components --- HSS, MME, and SPGW --- run in the CloudLab environment in the Utah lab~\cite{Duplyakin+:ATC19}. The CloudLab and UTD Lab are connected via Internet2. Due to the firewall restrictions in the UTD campus network, connectivity between the UTD Lab and the CloudLab environment is established through OpenVPN. Compute nodes in the CloudLab testbed are co-located and can therefore be used to test
virtualized EPC component migration within the same datacenter.
Both the VM and Docker Container components run in the CloudLab compute nodes using the same procedures already described for the OpenROADM testbed scenario, with the only exception that in the CloudLab testbed there is no optical transport (backhaul) network and OpenVPN service runs on all the core network VNF. The CloudLab testbed provides the opportunity to validate and test the robustness of the proposed and developed custom software in Section~\ref{sec:dc} --- software configuration changes, the new utility program, and CRIU code changes --- in an open environment outside the UTD in-house lab setting. 

\subsection{System Configuration} 
\label{SysConfig}

The system configuration details of both the OpenROADM and CloudLab testbed are reported in Table~\ref{tab:SystemConfig}. 
Compute nodes are Intel Xeon model in both testbeds.
However, the underlying compute hardware in the CloudLab testbed is 
more advanced and performing when compared to the one used in the OpenROADM testbed.

\begin{table}[htbp]
    \centering
    \caption{System configuration details.}
    \label{tab:SystemConfig}
    \begin{tabular}{|p{0.2\columnwidth}|p{0.32\columnwidth}|p{0.32\columnwidth}|} \hline
     \textbf{Description} & \textbf{OpenROADM} & \textbf{CloudLab} \\\hline
        Nodes & 1 control node, 1 network manager, and 8 compute nodes on each rack & 1 control node and 4 compute nodes on the same rack \\ \hline
        Product & Dell DCS8000Z & HP m510  \\ \hline
        CPU & 2 Intel Xeon E5-2680 processors@2.7GHz (16 cores, 2 threads/core) & 1 Intel Xeon D-1548 processors@2.00GHZ (8 cores, 2 threads/core)  \\ \hline
        Intel Architecture & Sandy Bridge & Broadwell \\ \hline
        Memory & RAM: 32GB, Disk: 256GB Flash Storage & RAM: 64GB, Disk: 256GB Flash Storage \\ \hline
        OpenStack-Management N/W & 1G-10G (Flexponder)-1G & 10G (Dual-port Mellanox ConnectX-3)  \\ \hline
        OpenStack-Tenant N/W & 40G-100G(Transponder)-40G & 10G (Dual-port Mellanox ConnectX-3)  \\ \hline
        QEMU version & 3.1.0 & 3.1.0  \\ \hline
        Libvirt version & 5.0.0 & 5.0.0  \\ \hline
        CRIU version & 3.12 & 3.12 \\ \hline
        Avg. CPU Utilization & < 10 \% & < 10 \%  \\ \hline
     \end{tabular}
    \vspace{-3mm}
\end{table}
In the OpenStack VM, the communication between the services is carried in the management network through advanced message queuing protocol (AMQP). Since the OpenStack VM migration traffic flows through the management network, the key parameter of interest for the VM migration is the management network data rate, which is relatively better in the CloudLab
as the compute nodes involved in the live migration procedure are all co-located. For the Container migration, since the Checkpoint and Restore services are done in the individual node, the parameters of interest are the better hardware architecture, cache size and the number of cores used. The same versions for the libvirt, KVM/QEMU, and CRIU packages are installed in both testbeds.

\subsection{Additional Observations}

The migration completion time is affected by a few key system parameters, which must be taken into account carefully.
First, flavors of computing instances (compute, memory, and storage capacity) may affect the time that is required to migrate both VMs and Containers to a new host. Table~\ref{tab:flavors} reports the flavors that are applied in this study. Second, the backhaul network round trip time may affect the completion time of the live migration.
The backhaul round trip time is affected by the Ethernet switch latency, optical transponder and switchponder latency, 
and finally optical signal propagation time across the network fiber. 
To test the effect of the signal propagation time on the migration completion time, multiple experiments are carried 
out while varying the route of the lightpath that is established between 
the two compute sites (racks) in the OpenROADM testbed (Fig.~\ref{fig:OpenROADM_Testbed}). The lightpath length is set to be a few meters, 25~km, and 50~km, respectively. Third, the use of OpenVPN may affect the migration completion time too. While OpenVPN must be used for the reasons discussed in previous sections --- e.g., overcome the OVS firewall driver configured in Neutron that blocks SCTP packets ---
some of the virtualized EPC components may still use floating IP connections in the OpenROADM testbed. To estimate the effect of these network interfaces on the migration completion time two configurations are investigated. In the \emph{OpenVPN} configuration all EPC components make use of OpenVPN. In the \emph{Floating IP} configuration the HSS and SPGW components make use of floating IP, while OpenVPN is still applied to the CU-MME SCTP connection.

\begin{table}[htb]
    \centering
    \caption{OpenStack flavors for the experimentation}
    \label{tab:flavors}
    \begin{tabular}{|p{0.1\columnwidth}|p{0.2\columnwidth}|p{0.1\columnwidth}|p{0.16\columnwidth}|p{0.15\columnwidth}|} \hline
     \textbf{No} & \textbf{Flavor Name} & \textbf{vCPUs} &\textbf{RAM [MB]} & \textbf{Disk [GB]} \\\hline
        1 & Small & 1 & 2048 & 20 \\ \hline
        2 & Medium & 2 & 4096 & 40  \\ \hline
    \end{tabular}
    \vspace{-1mm}
\end{table}

\section{Result Analysis}
\vspace{-0.05cm}

In this section performance indicators related to the migration procedure are first defined,
followed by the live migration analysis of the core network virtualized components. 
The performance indicators are evaluated and discussed for the following scenarios:
i) the OpenROADM testbed (OpenVPN) with varying lightpath lengths and flavor types; 
ii) the CloudLab-Federated testbed (OpenVPN); and 
iii) the use of Floating IP instead of OpenVPN in the OpenROADM testbed.

Migration of the chosen EPC component is performed once the UE connectivity 
is established and UE data is being transmitted and received over the RAN uplink and downlink.
Each experiment is repeated five times and the average is reported for each performance indicator
to mitigate stochastic variations of network, I/O, and process delays. 
When migrating VMs, it was noted that when the VNF is running
for a long period of time, its internal data storage increases (in terms of MB) due to the application collected logs. 
This data storage increase may cause variation of the network data traffic generated during the VM migration.

\subsection{Performance Indicators}

The main performance indicators considered in this study are \emph{migration time}, \emph{downtime}, \emph{network load} at the time of migration, and the \emph{UE service recovery time}~\cite{PerfMetrics, ContPerf, ContPerf1}.
Their definitions are given next.

\begin{itemize}

\item \textit{Migration time} is defined as the amount of time required to migrate a VNF from one host to another host. In the VM migration experiment, the migration time is the sum of the \emph{Pre-live}, \emph{Live} and \emph{Post-live} execution times, evaluated from the Nova log file. In the Container migration experiment, the migration time is the sum of the time intervals that are required to freeze the process, dump and transfer the metadata to the destination host, and restore the process at the destination host. The evaluation is carried out in millisecond resolution with the Shell Script that automates the migration process.

\item \textit{Downtime} is defined as the amount of time the VNF functionality is paused and unavailable. In the VM migration experiment, the downtime is associated to the execution of the final dirty page copy and the reconfiguration of the virtual interface bridge connectivity with the port settings at the destination host. Due to the chosen CRIU migration type, the downtime in the Container migration experiment is the same as the migration time. For both the VM and the Container, the downtime is measured using the ICMP ping of the VNF IP in millisecond resolution.

\item \textit{Network load} is defined as the amount of data transferred from the source host to the destination host.
In the VM migration experiment, the network load accounts for the transfer of the CPU state, memory state, network state, and disk state and it is measured using the network bandwidth monitoring tool named "iftop". For the Container migration experiment, the network load accounts for the process tree, the CPU state, memory page, namespace, and control group (cgroup) information and it is the size of metadata files checkpointed.

\item \textit{UE service recovery time} is defined as the time that is required to regain UE connectivity from the moment the UE is temporarily disconnected from the mobile network due to the migration of one of the EPC components. 
When using the OAI software modules the migration of either HSS or MME does not result in UE connectivity interruption. 
However, during the migration of SPGW, the UE uplink and downlink are temporarily disrupted and the UE service recovery time needs to be assessed. For both the VM and the Container, the UE Service Recovery Time is measured using the ICMP ping from the UE \emph{gtp} interface IP to the SPGW \emph{gtp} interface IP with one hundred milliseconds resolution.
\end{itemize}

\subsection{Migration Analysis of the HSS Component}

\begin{figure}[htbp]
     \vspace{-1.1em}
     \centering
     \includegraphics[trim=0 0 0 -1cm,width=1.0\columnwidth]{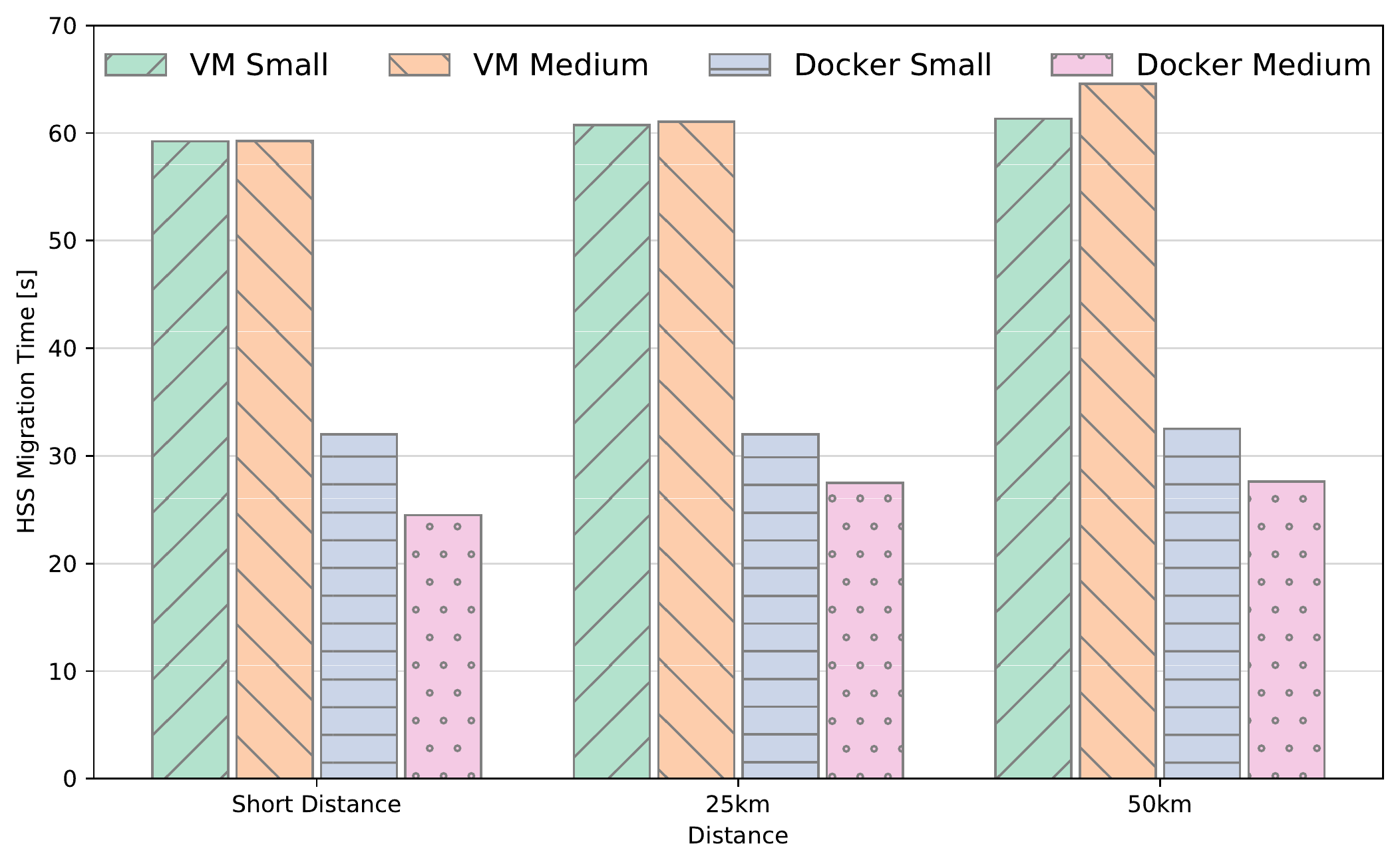}
    \caption{HSS - VM and Container migration time for three lightpath lengths and two flavor types.}
    \label{fig:HSS-MigTime}
    \vspace{-1.2em}
    \vspace{1em}
\end{figure}

Fig.~\ref{fig:HSS-MigTime} reports the migration time of both VM and Container running a virtualized HSS instance in the OpenROADM testbed for three lengths of the lightpath connecting the two racks (compute sites) and the two image flavor types in Table~\ref{tab:flavors}, respectively.

%
The migration time of the HSS VM is almost double that of the HSS Container regardless of the lightpath length and flavor type.
The extra time required by the VM migration is due to the VM disk image that must be migrated along with the memory page.

Only a modest extra time is required to complete all four migration types when using a longer lightpath, thus proving that these solutions can scale geographically. VM and Container migration times are differently affected by the flavor type.
The VM Medium flavor requires a modest extra migration time (more noticeable when using the 50~km lightpath)
compared to the VM Small flavor because of its increased image size. 
Transferring a larger image from the source host to the destination host takes extra time 
(magnified when the network round trip time is large).
Conversely, the Container Medium flavor requires less migration time (more noticeable when using the short lightpath)
compared to the Container Small flavor as its CPU core configuration enables the CRIU software to expedite both 
checkpoint and restoration executions.

\begin{figure}[htbp]
     \vspace{-1.1em}
     \centering
    \includegraphics[width=1.0\columnwidth]{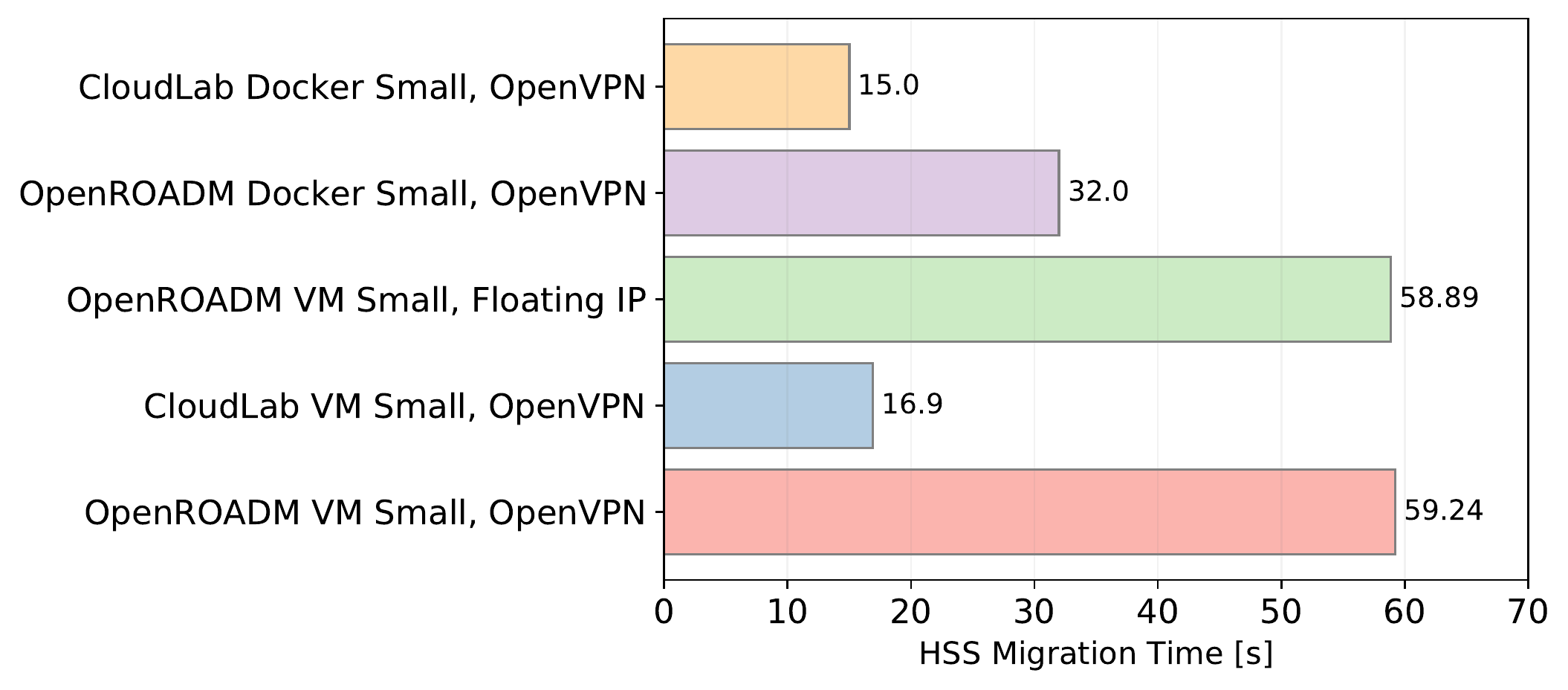} 
    \caption{Comparison of HSS migration time in the OpenROADM and CloudLab testbeds.}
    \label{fig:HSS-CloudMigTime}
    \vspace{-1.2em}
    \vspace{1em}
\end{figure}

Fig.~\ref{fig:HSS-CloudMigTime} compares the HSS migration times collected using both the OpenROADM
(lightpath length of few meters) and CloudLab testbed.
The latter testbed offers shorter migration times due to two system factors.
First, as specified in Sec.~\ref{SysConfig}, the management network in the CloudLab testbed operates at 10G 
compared to that of 1G-10G-1G used in the OpenROADM configuration.
The faster management network helps reduce the migration time in the VM-based experiments.
Second, both checkpoint and restoration executions in the CloudLab testbed are more performing due to
the Broadwell architecture running in the Intel Xeon servers with better L1 and L2 cache sizes
when compared to the Sandy Bridge architecture used in the OpenROADM testbed.
The more performing compute architecture of the CloudLab testbed helps reduce the migration time in the Container-based experiments.
In addition, the migration time is reduced a bit in the OpenROADM testbed, when 
the HSS application is configured with Floating IP instead of OpenVPN. 
By not using the OpenVPN client service package the 
VM image size is reduced, which in turn reduces its migration time.

Tables~\ref{Tbl:HSS-VMMigration-Breakup} and~\ref{tbl:HSS-ContMigration-Breakup} show 
the time taken by each phase of the VM and Container migration, respectively.
For the VM, the network loads for the HSS Small 
and Medium flavors at the time of migration are 3.47~GB and 3.68~GB, respectively.
For the HSS Docker Container, the metadata size is 173~MB regardless of the flavor type used. 

\begin{table}[]
\centering
\vspace{-0.3cm}
\caption{HSS - VM migration time breakdown}
\begin{adjustbox}{max width=\columnwidth}
\setlength{\tabcolsep}{3pt}
\setlength\arrayrulewidth{0.6pt}\arrayrulecolor{black}
\renewcommand{\arraystretch}{1.6}
\begin{tabular}{*{9}{|c}|}
\hline
{\textbf{}} & \multicolumn{7}{c|}{\textbf{OpenROADM}} & \multicolumn{1}{c|}{\textbf{CloudLab}} \\  
\hhline{~--------}
\multicolumn{1}{|c|}{\textbf{Configuration}} & 
\multicolumn{3}{c|}{\textbf{\cellcolor{lime!20}Flavor Small}} & \multicolumn{1}{c|}{\textbf{\cellcolor{brown!50}Flavor Small}} & \multicolumn{3}{c|} {\textbf{\cellcolor{yellow!15}Flavor Medium}} & \multicolumn{1}{c|}{\textbf{\cellcolor{gray!50}Flavor Small }}\\
\multicolumn{1}{|c|}{\textbf{[s]}} & 
\multicolumn{3}{c|}{\textbf{\cellcolor{lime!20}OpenVPN}} & \multicolumn{1}{c|}{\textbf{\cellcolor{brown!50}Floating IP}} & \multicolumn{3}{c|}{\textbf{\cellcolor{yellow!15}Floating IP}} & \multicolumn{1}{c|}{\textbf{\cellcolor{gray!50}OpenVPN }}\\[-1pt]
\hhline{~-------~}
 & \cellcolor{lime!20}Short & \cellcolor{lime!20}25km & \cellcolor{lime!20}50km & \cellcolor{brown!50}Short  & \cellcolor{yellow!15}Short & \cellcolor{yellow!15}25km & \cellcolor{yellow!15}50km & \cellcolor{gray!50}\\[-1pt]
 & \cellcolor{lime!20}Distance & \cellcolor{lime!20}& \cellcolor{lime!20}& \cellcolor{brown!50}Distance & \cellcolor{yellow!15}Distance & \cellcolor{yellow!15} &  \cellcolor{yellow!15} & \cellcolor{gray!50}\\ \hline
Pre-live & \cellcolor{lime!20}3.24 & \cellcolor{lime!20}3.25 & \cellcolor{lime!20}3.33 & \cellcolor{brown!50}3.29 & \cellcolor{yellow!15}3.26 & \cellcolor{yellow!15}3.53 & \cellcolor{yellow!15}3.56 & \cellcolor{gray!50}2.9 \\ \hline
Live & \cellcolor{lime!20}51 & \cellcolor{lime!20}51.5 & \cellcolor{lime!20}52.8 & \cellcolor{brown!50}50.3 & \cellcolor{yellow!15}51 & \cellcolor{yellow!15}52.52 & \cellcolor{yellow!15}55.8 & \cellcolor{gray!50}10 \\ \hline
Post-live & \cellcolor{lime!20}5 & \cellcolor{lime!20}5 & \cellcolor{lime!20}5.2 & \cellcolor{brown!50}5.3 & \cellcolor{yellow!15}5 & \cellcolor{yellow!15}5 & \cellcolor{yellow!15}5.2 & \cellcolor{gray!50}4 \\ \hline
\end{tabular}
\end{adjustbox}
\label{Tbl:HSS-VMMigration-Breakup}
\end{table}

\begin{table}[]
\centering
\caption{HSS - Container migration time breakdown}
\begin{adjustbox}{max width=\columnwidth}
\setlength{\tabcolsep}{3pt}
\setlength\arrayrulewidth{0.6pt}\arrayrulecolor{black}
\renewcommand{\arraystretch}{1.6}
\begin{tabular}{*{8}{|c}|}
\hline
\multicolumn{1}{|c|}{\textbf{Migration}} & \multicolumn{6}{c|}{\textbf{OpenROADM}} &
\multicolumn{1}{c|}{\textbf{CloudLab}} \\ [-1pt]
\hhline{~-------}
\multicolumn{1}{|c|}{\textbf{Metrics}} & \multicolumn{3}{c|}{\cellcolor{blue!25}\textbf{Flavor Small}} & \multicolumn{3}{c|}{\cellcolor{pink!50}\textbf{Flavor Medium}} & \multicolumn{1}{c|}{\cellcolor{gray!60}\textbf{Flavor Small}} \\[-1pt]
\multicolumn{1}{|c|}{\textbf{[s]}} & 
\multicolumn{3}{c|}{\textbf{\cellcolor{blue!25}OpenVPN}} &
\multicolumn{3}{c|} {\textbf{\cellcolor{pink!50}OpenVPN}} & \multicolumn{1}{c|}{\textbf{\cellcolor{gray!60}OpenVPN }} \\[-1pt]
\hhline{~------~}
{\textbf{}} & {\cellcolor{blue!25}Short} & \cellcolor{blue!25}25km & \cellcolor{blue!25}50km & \cellcolor{pink!50}Short & \cellcolor{pink!50}25km & \cellcolor{pink!50}50km & \cellcolor{gray!60}{} \\[-1pt]
\textbf{} & {\cellcolor{blue!25}Distance} & {\cellcolor{blue!25}} & {\cellcolor{blue!25}} & \cellcolor{pink!50}Distance & {\cellcolor{pink!50}} & {\cellcolor{pink!50}} & \cellcolor{gray!60}{} \\ 
\hline
Total Migration Time & \cellcolor{blue!25}32 & \cellcolor{blue!25}32 & \cellcolor{blue!25}32.33 & \cellcolor{pink!50}24.5 & \cellcolor{pink!50}27.5 & \cellcolor{pink!50}27.6 & \cellcolor{gray!60}15 \\ \hline
Checkpoint Time & \cellcolor{blue!25}13 & \cellcolor{blue!25}13 & \cellcolor{blue!25}13 & \cellcolor{pink!50}10 & \cellcolor{pink!50}11 & \cellcolor{pink!50}11 & \cellcolor{gray!60}6 \\ \hline
Metadata & \cellcolor{blue!25}3 & \cellcolor{blue!25}3 & \cellcolor{blue!25}3.33 & \cellcolor{pink!50}2.5 & \cellcolor{pink!50}2.5 & \cellcolor{pink!50}2.6 & \cellcolor{gray!60}2 \\ \hline
Restore Time & \cellcolor{blue!25}16 & \cellcolor{blue!25}16 & \cellcolor{blue!25}16 & \cellcolor{pink!50}12 & \cellcolor{pink!50}14 & \cellcolor{pink!50}14 & \cellcolor{gray!60}7 \\ \hline
\end{tabular}
\end{adjustbox}
\label{tbl:HSS-ContMigration-Breakup}
\end{table}
\begin{figure}[htbp]
     \vspace{-1.8em}
     \centering
     \includegraphics[trim=0 0 0 -1cm,width=0.9\columnwidth]{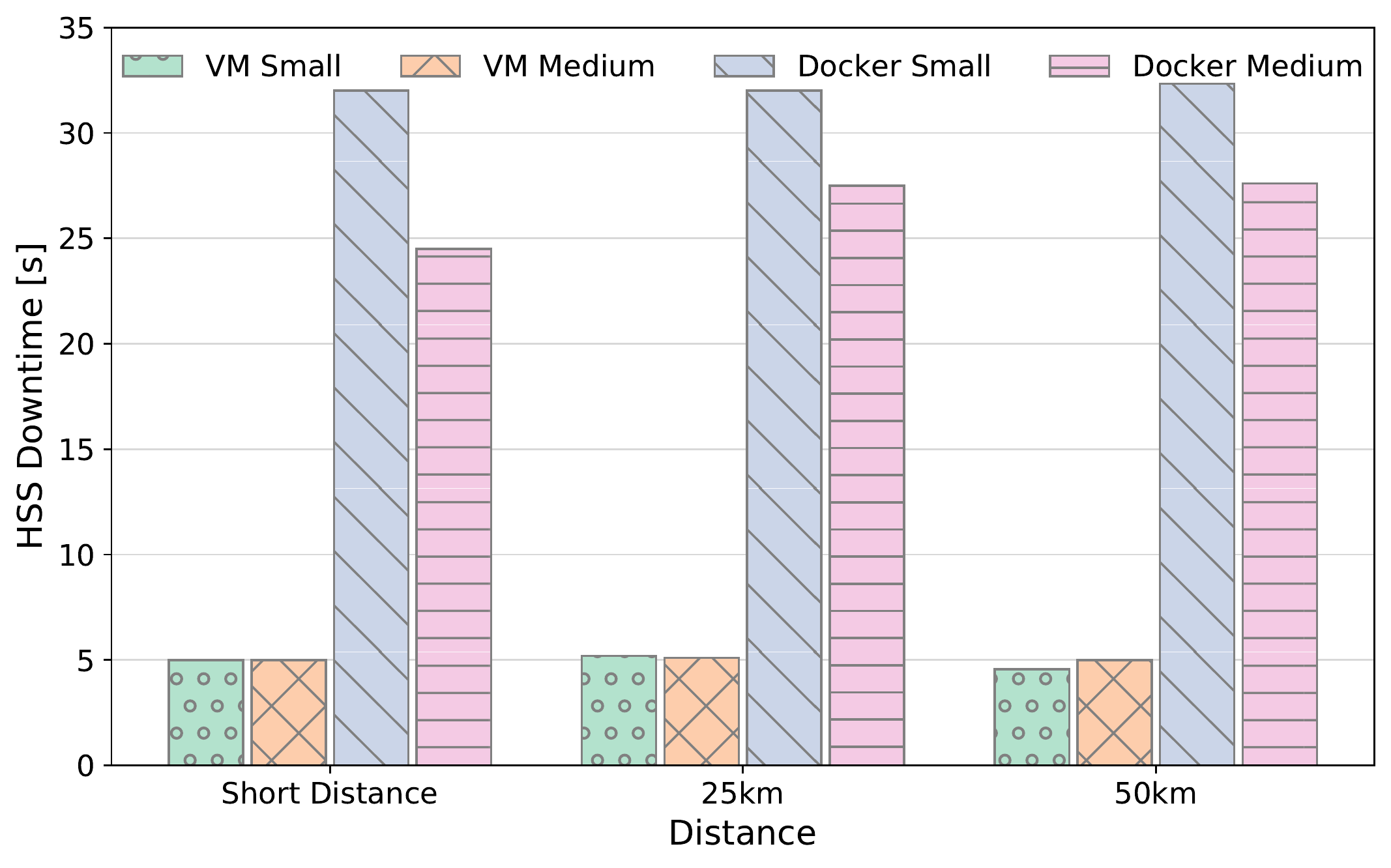}
    \caption{HSS - VM and Container downtime for three lightpath lengths and two flavor types.}
    \label{fig:HSS-DownTime}
    \vspace{-1.2em}
    \vspace{1em}
\end{figure}
As shown in Fig.~\ref{fig:HSS-DownTime} --- and in contrast to the previously presented migration time analysis ---
the downtime value for the HSS servicing Container is higher than that of the VM. 
The HSS VM downtime is mainly due to the virtual interface bridge and port reconfiguration. 
Other than that, the VNF service is not additionally paused during the VM live migration process. 
On the contrary, the HSS application running in the Container is paused once the checkpoint is initiated
and it resumes only after restoration is complete at the destination host.
Both lightpath length and flavor type do not have any significant impact on the HSS VM downtime. 
However, the flavor size impacts the HSS Container downtime as already noted for the migration time of the same experiments.

\begin{figure}[htbp]
     \vspace{-1.1em}
     \centering
     \includegraphics[trim=0 0 0 -1cm,width=1.0\columnwidth]{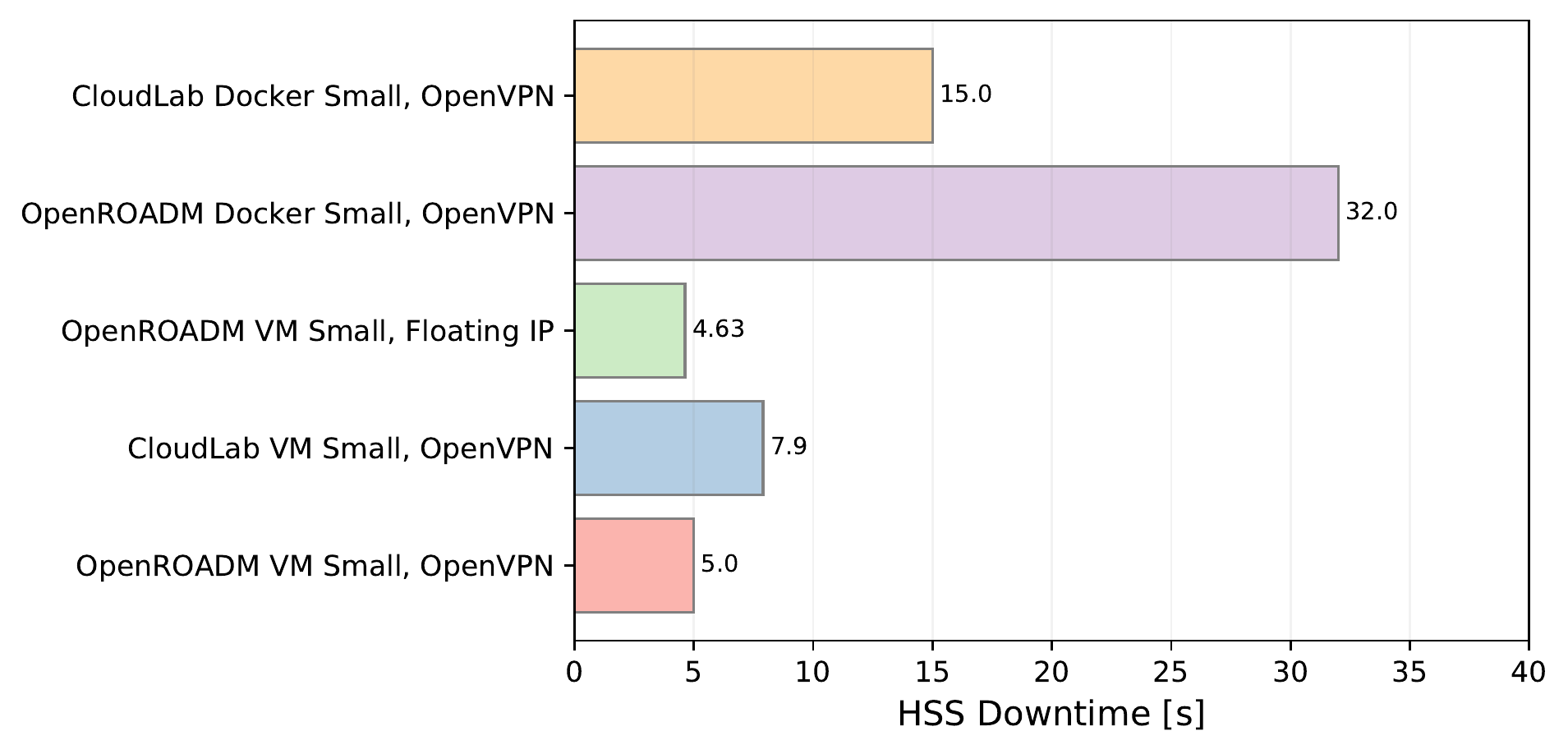}
    \caption{Comparison of HSS downtime in the OpenROADM and CloudLab testbeds}
    \label{fig:HSS-CloudDownTime}
    \vspace{1em}
\end{figure}

Fig.~\ref{fig:HSS-CloudDownTime} confirms that the downtime of the Docker Container in the CloudLab testbed is shorter
than that in the OpenROADM testbed, as previously noted for the migration times of these experiments.
As already mentioned, this outcome is mainly due to the superior processor 
architecture and cache level of the CloudLab testbed. 
More interestingly, the VM downtime is longer in the CloudLab testbed compared to the OpenROADM testbed.
The significant geographical distance between the CloudLab in Utah --- hosting the core network components ---
and the UTD Lab in Texas --- hosting the RAN components --- slows down the update procedures
for the Reverse Address Resolution Protocol (RARP) to determine the new host's IP address and reroute the OpenVPN client traffic.
Additionally, when Floating IP is used to replace OpenVPN for the HSS connectivity in the OpenROADM testbed, 
there is a slight reduction of the HSS downtime because in this configuration OpenVPN does not need
to establish a new route for the \emph{client-to-client} communication.

For an already attached UE, the mobile network service is not impacted by the temporary pause of the HSS component. 
However, any new UE attempting to perform attachment to the mobile network during the HSS migration procedure
would be affected by the HSS downtime.

\subsection{Migration Analysis of the MME Component}

\begin{figure}[htbp]
     \centering
     \includegraphics[trim=0 0 0 -1cm,width=1.0\columnwidth]{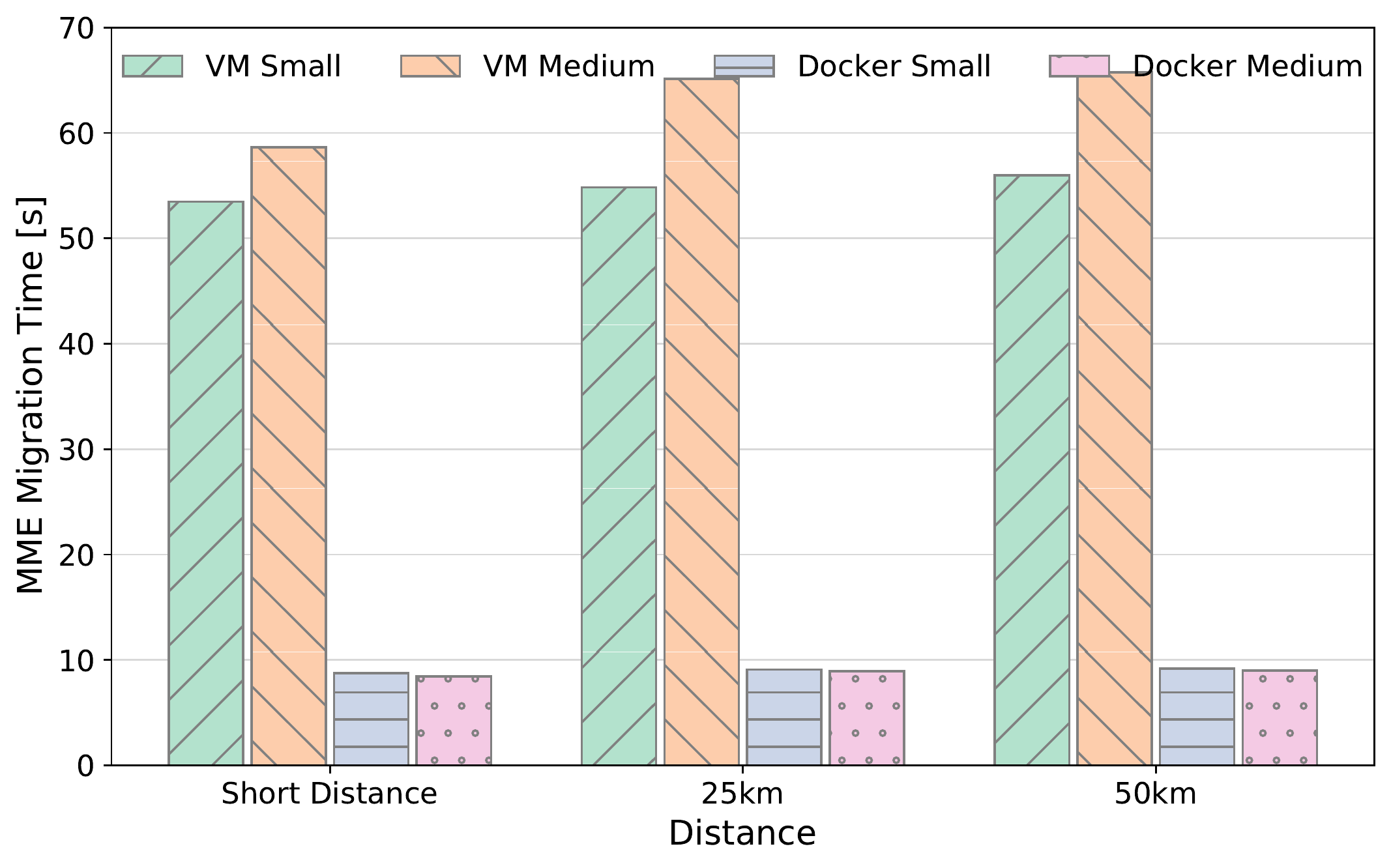}
    \caption{MME - VM and Container migration time for three lightpath lengths and two flavor types.}
    \label{fig:MME-MigTime}
    \vspace{-1.2em}
    \vspace{1em}
\end{figure}
\begin{figure}[htbp]
    \vspace{-1.8em}
     \centering
    \includegraphics[trim=0 0 0 -1cm,width=1.0\columnwidth]{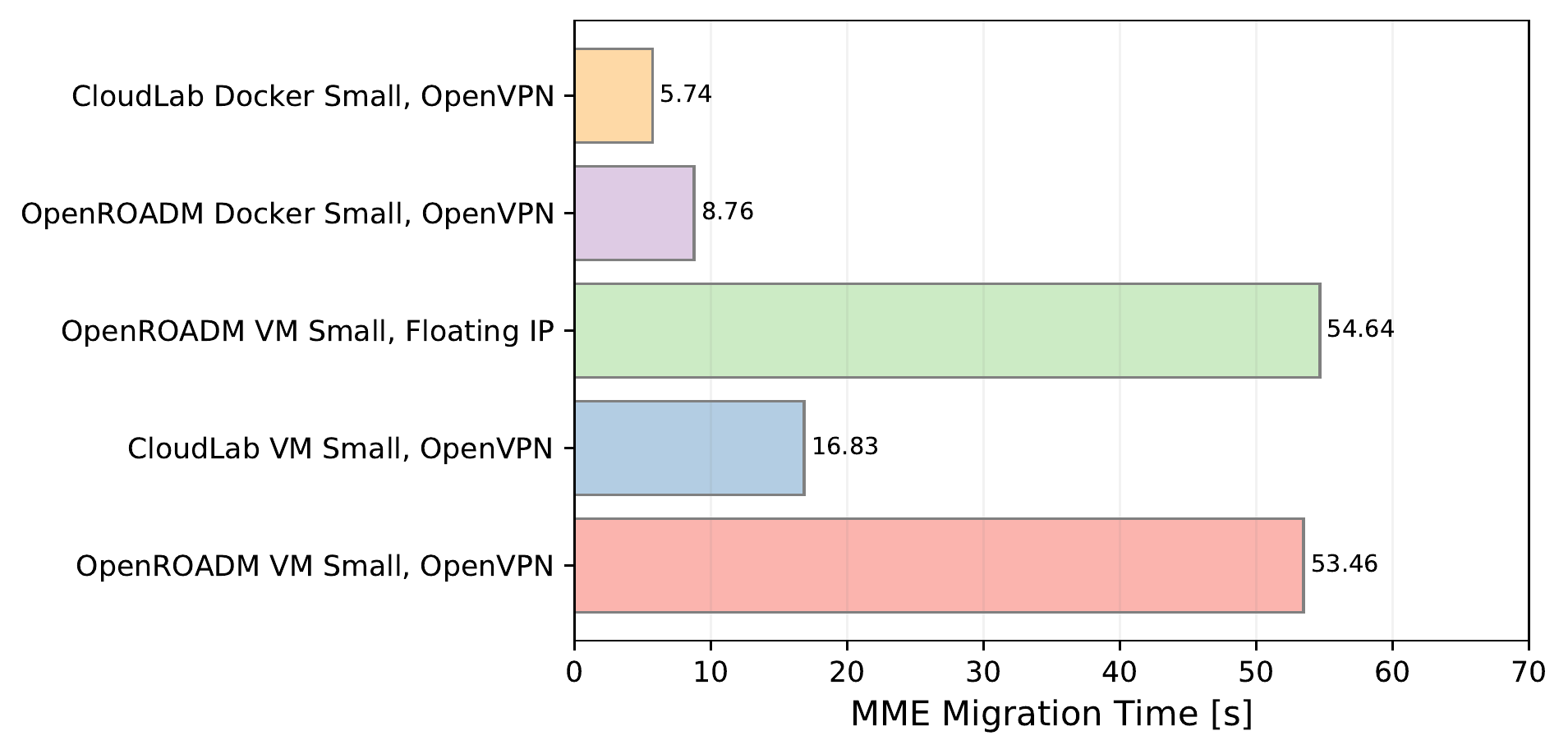}
    \caption{Comparison of MME migration time in the OpenROADM and CloudLab testbeds.}
    \label{fig:MME-CloudMigrationTime}
    \vspace{-1.2em}
    \vspace{1em}
\end{figure}

As reported in Fig.~\ref{fig:MME-MigTime}, the migration time of the MME VM compared to the MME Container 
is six and seven times larger when using the Small and Medium flavor, respectively.
The MME VM network loads for the Small and Medium flavors at the time of migration are 3.53~GB and 4.16~GB, respectively. 
Both larger image size and longer lightpath length tend to increase the migration time of the MME VM.
In comparison, the Docker Container metadata size is 42~MB in 
the Small flavor and only grows by 0.3~MB in the Medium flavor.
Migration time of the MME Container is almost unaffected by the flavor type and 
lightpath length, with reported variations in the subsecond range. The MME migration times in the OpenROADM testbed and CloudLab testbed are reported in Fig.~\ref{fig:MME-CloudMigrationTime} for a few configurations. As already noted for the HSS migration, the superior compute architecture available in the CloudLab testbed achieves shorter MME migration times compared to those in the OpenROADM testbed. For the same reason, the VM migration time compared to the Container migration time is only about three times longer in the CloudLab testbed. For the MME VM Floating IP scenario, the MME configuration uses the Floating IP to communicate with the HSS and the SPGW software and the OpenVPN IP to communicate with the CU software (to avoid SCTP packets blocking at the OVS).
The migration time of the Small flavor VM in the OpenROADM testbed is 
not significantly affected by the use of Floating IP in place of OpenVPN. This outcome is not surprising as the Floating IP network configuration can only be applied to the HSS and SPGW component and cannot be applied completely to the MME component that requires SCTP to run over OpenVPN to overcome the OVS firewall in Neutron.


Tables~\ref{Tbl:MME-VMMigration-Breakup} and~\ref{tbl:MME-ContMigration-Breakup} report
the time taken during each phase of the VM and the Container migration of the MME component. 
The impact due to lightpath length and flavor type variations is more noticeable in 
the \emph{Live} phase (disk and memory copy phase) of the VM migration. 
For the MME Container, lightpath length variations primarily affect the metadata transfer time 
whereas flavor change variations primarily affects the Checkpoint and Restore execution times.
%
%
\begin{table}[]
\centering
\vspace{-0.3cm}
\caption{MME - VM migration time breakdown}
\begin{adjustbox}{max width=\columnwidth}
\setlength{\tabcolsep}{3pt}
\setlength\arrayrulewidth{0.6pt}\arrayrulecolor{black}
\renewcommand{\arraystretch}{1.6}
\begin{tabular}{*{9}{|c}|}
\hline
{\textbf{}} & \multicolumn{7}{c|}{\textbf{OpenROADM}} & \multicolumn{1}{c|}{\textbf{CloudLab}} \\ [-1pt]
\hhline{~--------}
\multicolumn{1}{|c|}{\textbf{Configuration}} & 
\multicolumn{3}{c|}{\textbf{\cellcolor{lime!20}Flavor Small}} & \multicolumn{1}{c|}{\textbf{\cellcolor{brown!50}Flavor Small}} & \multicolumn{3}{c|} {\textbf{\cellcolor{yellow!15}Flavor Medium}} & \multicolumn{1}{c|}{\textbf{\cellcolor{gray!50}Flavor Small }}\\[-1pt]
\multicolumn{1}{|c|}{\textbf{[s]}} & 
\multicolumn{3}{c|}{\textbf{\cellcolor{lime!20}OpenVPN}} & \multicolumn{1}{c|}{\textbf{\cellcolor{brown!50}Floating IP}} & \multicolumn{3}{c|} {\textbf{\cellcolor{yellow!15}Floating IP}} & \multicolumn{1}{c|}{\textbf{\cellcolor{gray!50}OpenVPN }} \\[-1pt]
\hhline{~-------~}
 & \cellcolor{lime!20}Short & \cellcolor{lime!20}25km & \cellcolor{lime!20}50km & \cellcolor{brown!50}Short  & \cellcolor{yellow!15}Short & \cellcolor{yellow!15}25km & \cellcolor{yellow!15}50km & \cellcolor{gray!50}\\[-1pt]
 & \cellcolor{lime!20}Distance & \cellcolor{lime!20}& \cellcolor{lime!20}& \cellcolor{brown!50}Distance & \cellcolor{yellow!15}Distance & \cellcolor{yellow!15} &  \cellcolor{yellow!15} & \cellcolor{gray!50}\\ \hline
Pre-live & \cellcolor{lime!20}3.26 & \cellcolor{lime!20}3.34 & \cellcolor{lime!20}3.38 & \cellcolor{brown!50}3.14 & \cellcolor{yellow!15}3.23 & \cellcolor{yellow!15}3.32 & \cellcolor{yellow!15}3.35 & \cellcolor{gray!50}2.83 \\ \hline
Live & \cellcolor{lime!20}45.2 & \cellcolor{lime!20}46 & \cellcolor{lime!20}46.8 & \cellcolor{brown!50}46 & \cellcolor{yellow!15}50 & \cellcolor{yellow!15}56 & \cellcolor{yellow!15}56.4 & \cellcolor{gray!50}10 \\ \hline
Post-live & \cellcolor{lime!20}5 & \cellcolor{lime!20}5.5 & \cellcolor{lime!20}5.8 & \cellcolor{brown!50}5.5 & \cellcolor{yellow!15}5.4 & \cellcolor{yellow!15}5.8 & \cellcolor{yellow!15}6 & \cellcolor{gray!50}4 \\ \hline
\end{tabular}
\end{adjustbox}
\label{Tbl:MME-VMMigration-Breakup}
\end{table}
\begin{table}[]
\centering
\caption{MME - Container migration time breakdown}
\begin{adjustbox}{max width=\columnwidth}
\setlength{\tabcolsep}{3pt}
\setlength\arrayrulewidth{0.6pt}\arrayrulecolor{black}
\renewcommand{\arraystretch}{1.6}
\begin{tabular}{*{8}{|c}|}
\hline
\multicolumn{1}{|c|}{\textbf{Migration}} & \multicolumn{6}{c|}{\textbf{OpenROADM}} & 
\multicolumn{1}{c|}{\textbf{CloudLab}} \\ [-1pt]
\hhline{~-------}
\multicolumn{1}{|c|}{\textbf{Metrics}} & \multicolumn{3}{c|}{\cellcolor{blue!25}\textbf{Flavor Small}} & \multicolumn{3}{c|}{\cellcolor{pink!50}\textbf{Flavor Medium}} & \multicolumn{1}{c|}{\cellcolor{gray!60}\textbf{Flavor Small}} \\ [-1pt]
\multicolumn{1}{|c|}{\textbf{[s]}} & 
\multicolumn{3}{c|}{\textbf{\cellcolor{blue!25}OpenVPN}} & 
\multicolumn{3}{c|} {\textbf{\cellcolor{pink!50}OpenVPN}} & \multicolumn{1}{c|}{\textbf{\cellcolor{gray!60}OpenVPN }} \\[-1pt]
\hhline{~------~}
{\textbf{}} & {\cellcolor{blue!25}Short} & \cellcolor{blue!25}25km & \cellcolor{blue!25}50km & \cellcolor{pink!50}Short & \cellcolor{pink!50}25km & \cellcolor{pink!50}50km & \cellcolor{gray!60}{} \\[-1pt]
\textbf{} & {\cellcolor{blue!25}Distance} & {\cellcolor{blue!25}} & {\cellcolor{blue!25}} & \cellcolor{pink!50}Distance & {\cellcolor{pink!50}} & {\cellcolor{pink!50}} & \cellcolor{gray!60}{} \\ 
\hline
Total Migration Time & \cellcolor{blue!25}8.76 & \cellcolor{blue!25}9.1 & \cellcolor{blue!25}9.19 & \cellcolor{pink!50}8.46 & \cellcolor{pink!50}8.94 & \cellcolor{pink!50}8.99 & \cellcolor{gray!60}5.74 \\ \hline
Checkpoint Time & \cellcolor{blue!25}4.2 & \cellcolor{blue!25}4.29 & \cellcolor{blue!25}4.34 & \cellcolor{pink!50}4.25 & \cellcolor{pink!50}4.47 & \cellcolor{pink!50}4.4 & \cellcolor{gray!60}2.19 \\ \hline
Metadata & \cellcolor{blue!25}1.16 & \cellcolor{blue!25}1.39 & \cellcolor{blue!25}1.43 & \cellcolor{pink!50}1.01 & \cellcolor{pink!50}1.22 & \cellcolor{pink!50}1.32 & \cellcolor{gray!60}1.16 \\ \hline
Restore Time & \cellcolor{blue!25}3.4 & \cellcolor{blue!25}3.42 & \cellcolor{blue!25}3.42 & \cellcolor{pink!50}3.2 & \cellcolor{pink!50}3.25 & \cellcolor{pink!50}3.27 & \cellcolor{gray!60}2.39 \\ \hline
\end{tabular}
\end{adjustbox}
\label{tbl:MME-ContMigration-Breakup}
\vspace{-0.3cm}
\end{table}
%
The Checkpoint and Restore execution times for the MME component are 
only a fraction of the corresponding execution times for the HSS component.
The cause for this difference is the metadata size which is smaller in the former component.
The MME application mainly handles the "control-plane socket connection information," stored in its metadata. In contrast, the HSS metadata stores the user database information and, as the number of UE client information increases, the HSS metadata size increases further. 

\begin{figure}[htbp]
     \vspace{-1.1em}
     \centering
    \includegraphics[trim=0 0 0 -1cm,width=1.0\columnwidth]{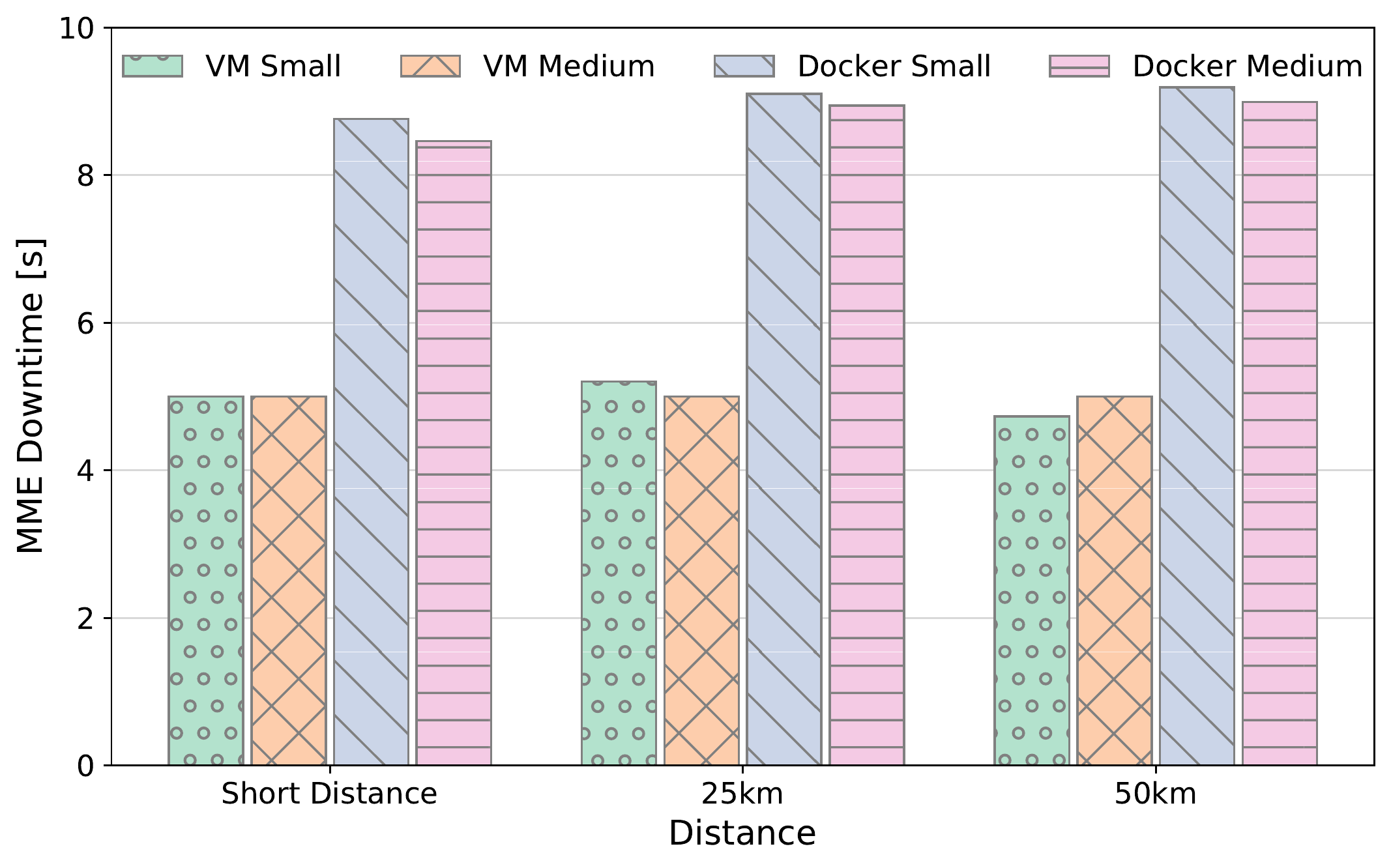}
    \caption{MME - VM and Container downtime for three lightpath lengths and two flavor types.}
    \label{fig:MME-DownTime}
    \vspace{-1.2em}
    \vspace{1em}
\end{figure}

Fig.~\ref{fig:MME-DownTime} shows that for the virtualized MME, the Container downtime is almost double the value of VM. 
As specified earlier, for the Docker Container, the MME service is paused at the Checkpoint initiation and it is resumed after the Restore procedure -- contributing the longer downtime value. As mentioned earlier, for the MME VM, primarily the virtual interface bridge and port reconfiguration influences the VM downtime.

\begin{figure}[htbp]
     \vspace{-1.1em}
     \centering
     \includegraphics[trim=0 0 0 -1cm,width=1.0\columnwidth]{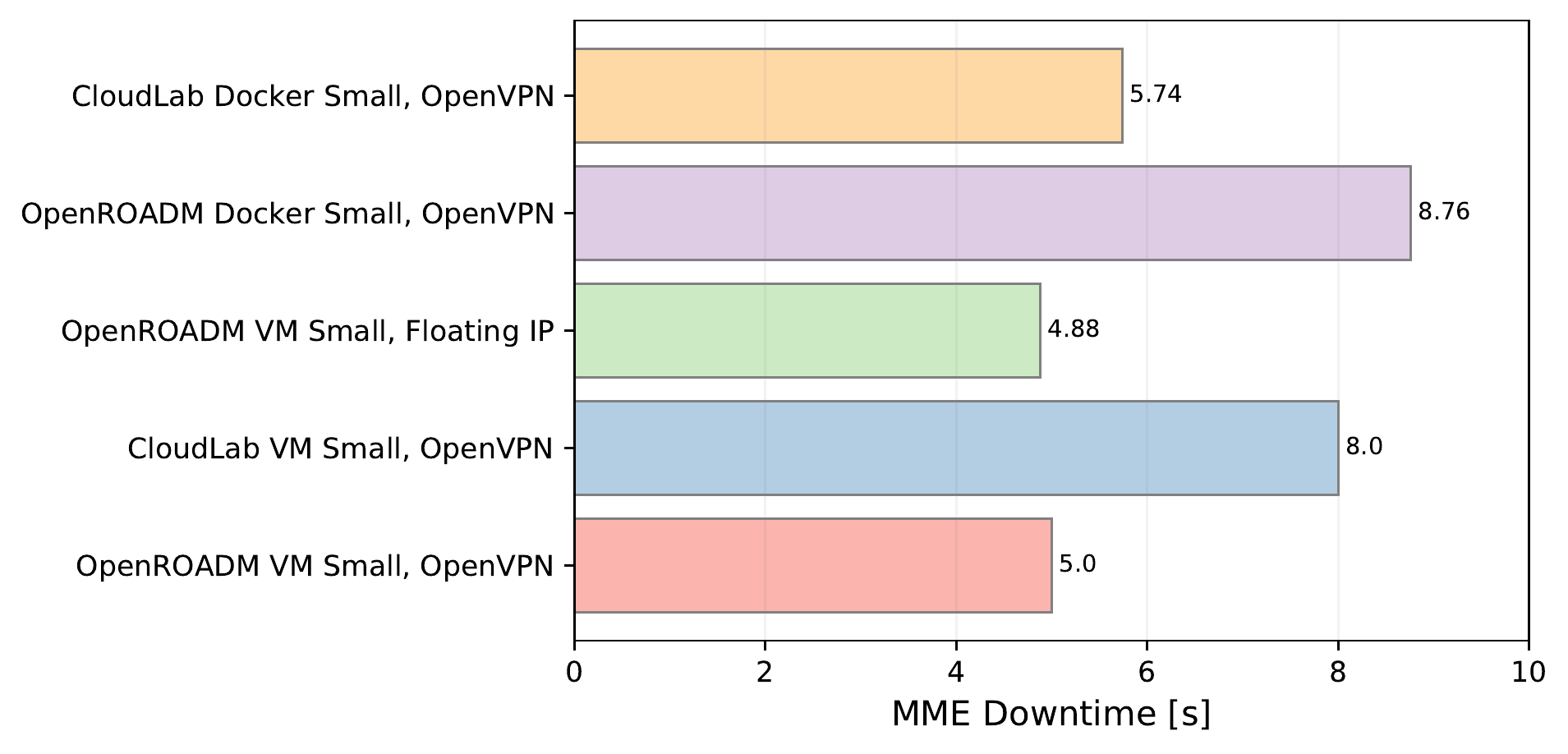}
    \caption{Comparison of MME downtime in the OpenROADM and CloudLab testbeds.}
    \label{fig:MME-CloudDownTime}
    \vspace{-1.2em}
    \vspace{1em}
\end{figure}
The OpenROADM testbed and the CloudLab testbed comparison confirms the influence of system configuration characteristics for tuning the migration parameters. As shown in Fig.~\ref{fig:MME-CloudDownTime}, the VM downtime is longer than the Container downtime in the CloudLab testbed, in contrast with the opposite trend observed in the OpenROADM testbed. 
The geographical location with the OpenVPN configuration is the prime factor behind this observation for the VM in the CloudLab. Considering the MME Docker Container service, the CloudLab with its better underlying hardware architecture shows the superiority in downtime than the OpenROADM Docker testbed. In the OpenROADM environment, the MME service downtime is reduced by a thin margin when the Floating IP is configured in the MME application instead of OpenVPN IP.

For the attached UE, the OAI based UE service is not disrupted by the MME migration as long as there is no handover or tracking area update related signaling functionality requirement. 

\subsection{Migration Analysis of the SPGW Component}

In this section --- in addition to the migration time and downtime of the SPGW component --- the end-user service interruption is visualized with the UE service recovery time performance indicator.

\begin{figure}[htbp]
     \vspace{-1.1em}
     \centering
     \includegraphics[trim=0 0 0 -1cm,width=1.0\columnwidth]{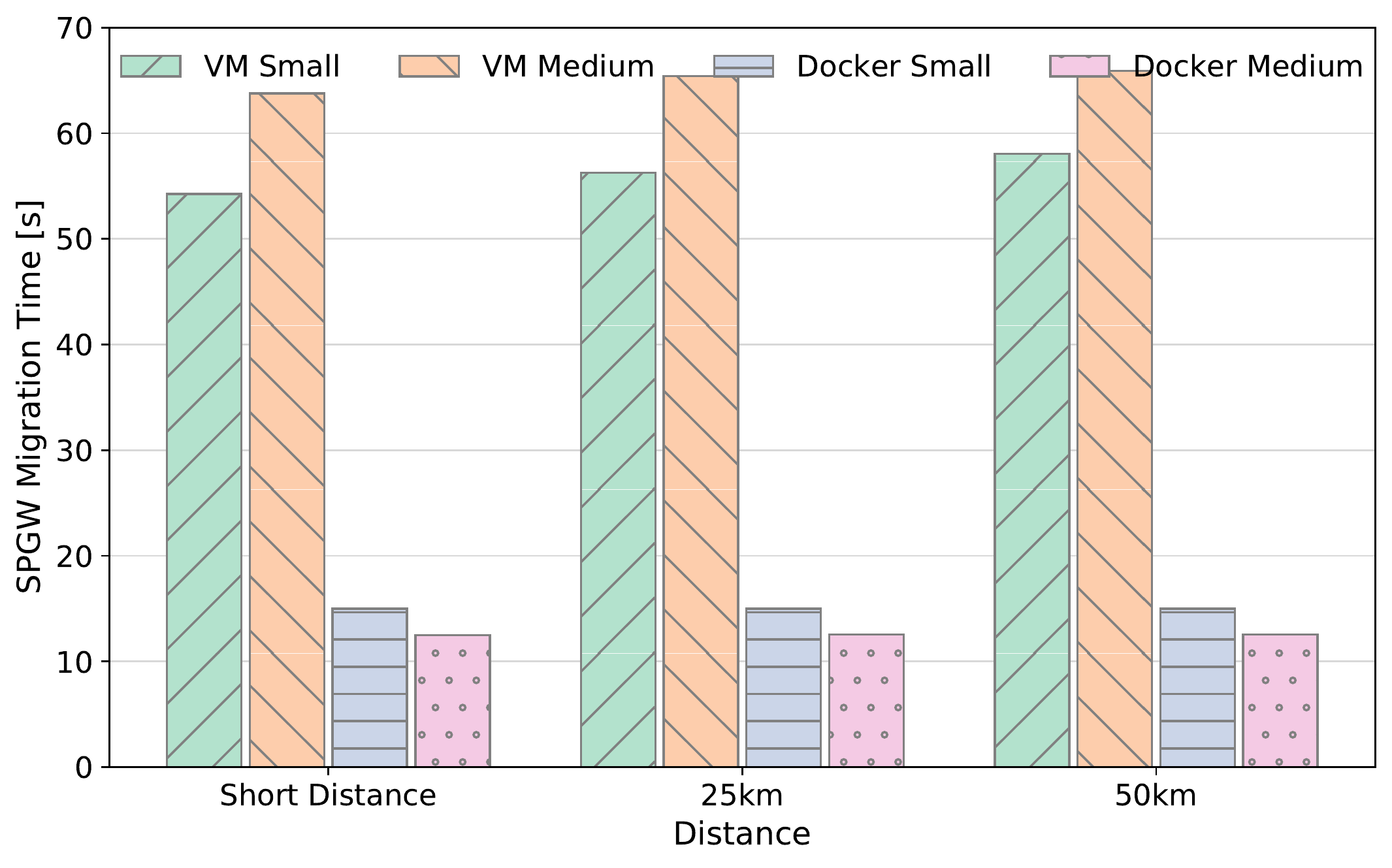}
    \caption{SPGW - VM and Container migration time for three lightpath lengths and two flavor types.}
    \label{fig:SPGW-MigTime}
    \vspace{-1.2em}
    \vspace{1em}
\end{figure}

Fig.~\ref{fig:SPGW-MigTime} presents the migration time of both the VM 
and Container running a virtualized SPGW in the OpenROADM testbed. 
The migration time of the Container is significantly less compared to that of the VM. 
For the VM, the image sizes for the SPGW Small and Medium flavors at the time of migration are 3.53~GB and 3.87~GB, respectively. 
For the SPGW Docker Container, the metadata size is in terms of 100~MB independent of the flavor type used.

\begin{figure}[htbp]
     \vspace{-1.1em}
     \centering
    \includegraphics[trim=0 0 0 -1cm,width=1.0\columnwidth]{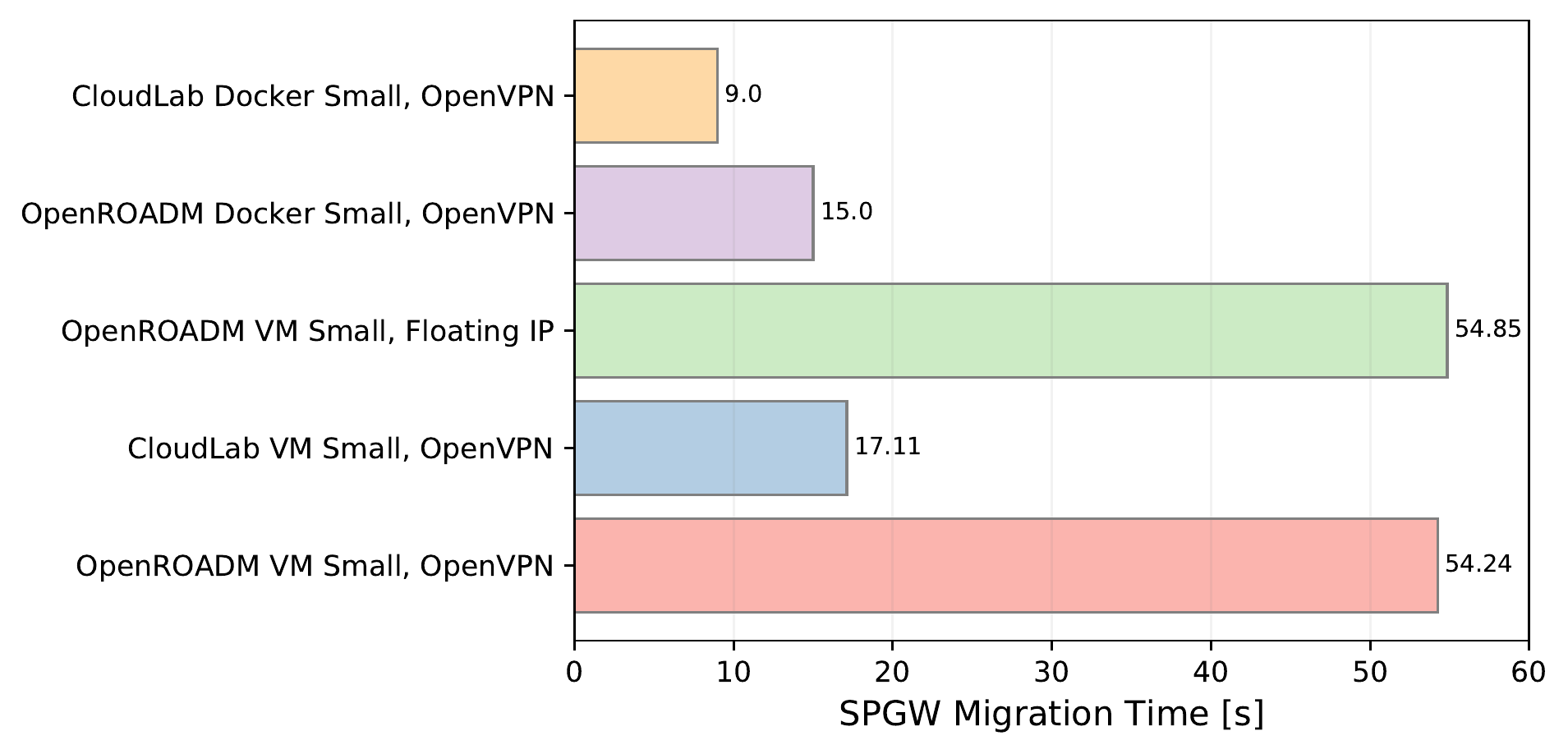} 
    \caption{Comparison of SPGW migration time in the OpenROADM and CloudLab testbeds.}
    \label{fig:SPGW-CloudMigrationTime}
    \vspace{-1.2em}
    \vspace{1em}
\end{figure}

Fig.~\ref{fig:SPGW-CloudMigrationTime} compares the SPGW migration time collected in both the OpenROADM and CloudLab testbed. 
This result confirms --- in the same way as HSS and MME migration time --- the CloudLab testbed migration time value is considerably less than the OpenROADM migration time due to the system factors. 
In addition, from the CloudLab results, the migration time of the SPGW VM is almost double than that of the SPGW Container. 
This is because of the extra time required to migrate the VM disk image (3.53~GB). 

Tables~\ref{Tbl:SPGW-VMMigration-Breakup} and~\ref{tbl:SPGW-ContMigration-Breakup} reports the time taken by each phase of the SPGW VM and Container migration. 
For the VM migration, the flavor size and the lightpath influence the \emph{Pre-live} and \emph{Live} migration phases, however, no change is observed for the \emph{Post-live} migration phase. 
For the Container migration, the Checkpoint and Restoration time for the SPGW application is relatively lower (metadata size 100~MB) than the HSS service (metadata size 173~MB) and higher than the MME service (metadata size 42~MB). 
This shows that the time to Checkpoint and Restore an application depends on various factors such as dumping the opened files, pages, core and task information, etc. 

\begin{table}[]
\centering
\vspace{-0.3cm}
\caption{SPGW - VM migration time breakdown}
\begin{adjustbox}{max width=\columnwidth}
\setlength{\tabcolsep}{3pt}
\setlength\arrayrulewidth{0.6pt}\arrayrulecolor{black}
\renewcommand{\arraystretch}{1.6}
\begin{tabular}{*{9}{|c}|}
\hline
{\textbf{}} & \multicolumn{7}{c|}{\textbf{OpenROADM}} & \multicolumn{1}{c|}{\textbf{CloudLab}} \\ [-1pt]
\hhline{~--------}
\multicolumn{1}{|c|}{\textbf{Configuration}} & 
\multicolumn{3}{c|}{\textbf{\cellcolor{lime!20}Flavor Small}} & \multicolumn{1}{c|}{\textbf{\cellcolor{brown!50}Flavor Small}} & \multicolumn{3}{c|} {\textbf{\cellcolor{yellow!15}Flavor Medium}} & \multicolumn{1}{c|}{\textbf{\cellcolor{gray!50}Flavor Small }}\\[-1pt]
\multicolumn{1}{|c|}{\textbf{[s]}} & 
\multicolumn{3}{c|}{\textbf{\cellcolor{lime!20}OpenVPN}} & \multicolumn{1}{c|}{\textbf{\cellcolor{brown!50}Floating IP}} & \multicolumn{3}{c|} {\textbf{\cellcolor{yellow!15}Floating IP}} & \multicolumn{1}{c|}{\textbf{\cellcolor{gray!50}OpenVPN}} \\[-1pt]
\hhline{~-------~}
 & \cellcolor{lime!20}Short & \cellcolor{lime!20}25km & \cellcolor{lime!20}50km & \cellcolor{brown!50}Short  & \cellcolor{yellow!15}Short & \cellcolor{yellow!15}25km & \cellcolor{yellow!15}50km & \cellcolor{gray!50}\\[-1pt]
 & \cellcolor{lime!20}Distance & \cellcolor{lime!20}& \cellcolor{lime!20}& \cellcolor{brown!50}Distance & \cellcolor{yellow!15}Distance & \cellcolor{yellow!15} &  \cellcolor{yellow!15} & \cellcolor{gray!50}\\ \hline
Pre-live & \cellcolor{lime!20}3.24 & \cellcolor{lime!20}3.25 & \cellcolor{lime!20}3.26 & \cellcolor{brown!50}3.25 & \cellcolor{yellow!15}3.26 & \cellcolor{yellow!15}3.41 & \cellcolor{yellow!15}3.51 & \cellcolor{gray!50}2.97 \\ \hline
Live & \cellcolor{lime!20}45 & \cellcolor{lime!20}47 & \cellcolor{lime!20}48.8 & \cellcolor{brown!50}45.6 & \cellcolor{yellow!15}54.5 & \cellcolor{yellow!15}56 & \cellcolor{yellow!15}56.4 & \cellcolor{gray!50}10.14 \\ \hline
Post-live & \cellcolor{lime!20}6 & \cellcolor{lime!20}6 & \cellcolor{lime!20}6 & \cellcolor{brown!50}6 & \cellcolor{yellow!15}6 & \cellcolor{yellow!15}6 & \cellcolor{yellow!15}6 & \cellcolor{gray!50}4 \\ \hline
\end{tabular}
\end{adjustbox}
\label{Tbl:SPGW-VMMigration-Breakup}
\end{table}
\begin{table}[]
\centering
\caption{SPGW - Container migration time breakdown}
\begin{adjustbox}{max width=\columnwidth}
\setlength{\tabcolsep}{3pt}
\setlength\arrayrulewidth{0.6pt}\arrayrulecolor{black}
\renewcommand{\arraystretch}{1.6}
\begin{tabular}{*{8}{|c}|}
\hline
\multicolumn{1}{|c|}{\textbf{Migration}} & \multicolumn{6}{c|}{\textbf{OpenROADM}} &
\multicolumn{1}{c|}{\textbf{CloudLab}} \\ [-1pt]
\hhline{~-------}
\multicolumn{1}{|c|}{\textbf{Metrics}} & \multicolumn{3}{c|}{\cellcolor{blue!25}\textbf{Flavor Small}} & \multicolumn{3}{c|}{\cellcolor{pink!50}\textbf{Flavor Medium}} & {\cellcolor{gray!60}\textbf{Flavor Small}} \\ [-1pt]
\multicolumn{1}{|c|}{\textbf{[s]}} & 
\multicolumn{3}{c|}{\textbf{\cellcolor{blue!25}OpenVPN}} &  \multicolumn{3}{c|} {\textbf{\cellcolor{pink!50}OpenVPN}} & \multicolumn{1}{c|}{\textbf{\cellcolor{gray!60}OpenVPN }} \\[-1pt]
\hhline{~------~}
{\textbf{}} & {\cellcolor{blue!25}Short} & \cellcolor{blue!25}25km & \cellcolor{blue!25}50km & \cellcolor{pink!50}Short & \cellcolor{pink!50}25km & \cellcolor{pink!50}50km & \cellcolor{gray!60}{} \\[-1pt]
\textbf{} & {\cellcolor{blue!25}Distance} & {\cellcolor{blue!25}} & {\cellcolor{blue!25}} & \cellcolor{pink!50}Distance & {\cellcolor{pink!50}} & {\cellcolor{pink!50}} & \cellcolor{gray!60}{} \\ 
\hline
Total Migration Time & \cellcolor{blue!25}15 & \cellcolor{blue!25}15 & \cellcolor{blue!25}15 & \cellcolor{pink!50}12.49 & \cellcolor{pink!50}12.56 & \cellcolor{pink!50}12.56 & \cellcolor{gray!60}9 \\ \hline
Checkpoint Time & \cellcolor{blue!25}7 & \cellcolor{blue!25}7 & \cellcolor{blue!25}7 & \cellcolor{pink!50}6 & \cellcolor{pink!50}6 & \cellcolor{pink!50}6 & \cellcolor{gray!60}4 \\ \hline
Metadata & \cellcolor{blue!25}2 & \cellcolor{blue!25}2 & \cellcolor{blue!25}2 & \cellcolor{pink!50}1.49 & \cellcolor{pink!50}1.56 & \cellcolor{pink!50}1.56 & \cellcolor{gray!60}1 \\ \hline
Restore Time & \cellcolor{blue!25}6 & \cellcolor{blue!25}6 & \cellcolor{blue!25}6 & \cellcolor{pink!50}5 & \cellcolor{pink!50}5 & \cellcolor{pink!50}5 & \cellcolor{gray!60}4 \\ \hline
UE SRT & \cellcolor{blue!25}2 & \cellcolor{blue!25}2 & \cellcolor{blue!25}2.24 & \cellcolor{pink!50}1.43 & \cellcolor{pink!50}1.55 & \cellcolor{pink!50}2 & \cellcolor{gray!60}2.28 \\ \hline
\end{tabular}
\end{adjustbox}
\label{tbl:SPGW-ContMigration-Breakup}
\vspace{-0.3cm}
\end{table}

\begin{figure}[htbp]
     \vspace{-1.1em}
     \centering
     \includegraphics[trim=0 0 0 -1cm,width=1.0\columnwidth]{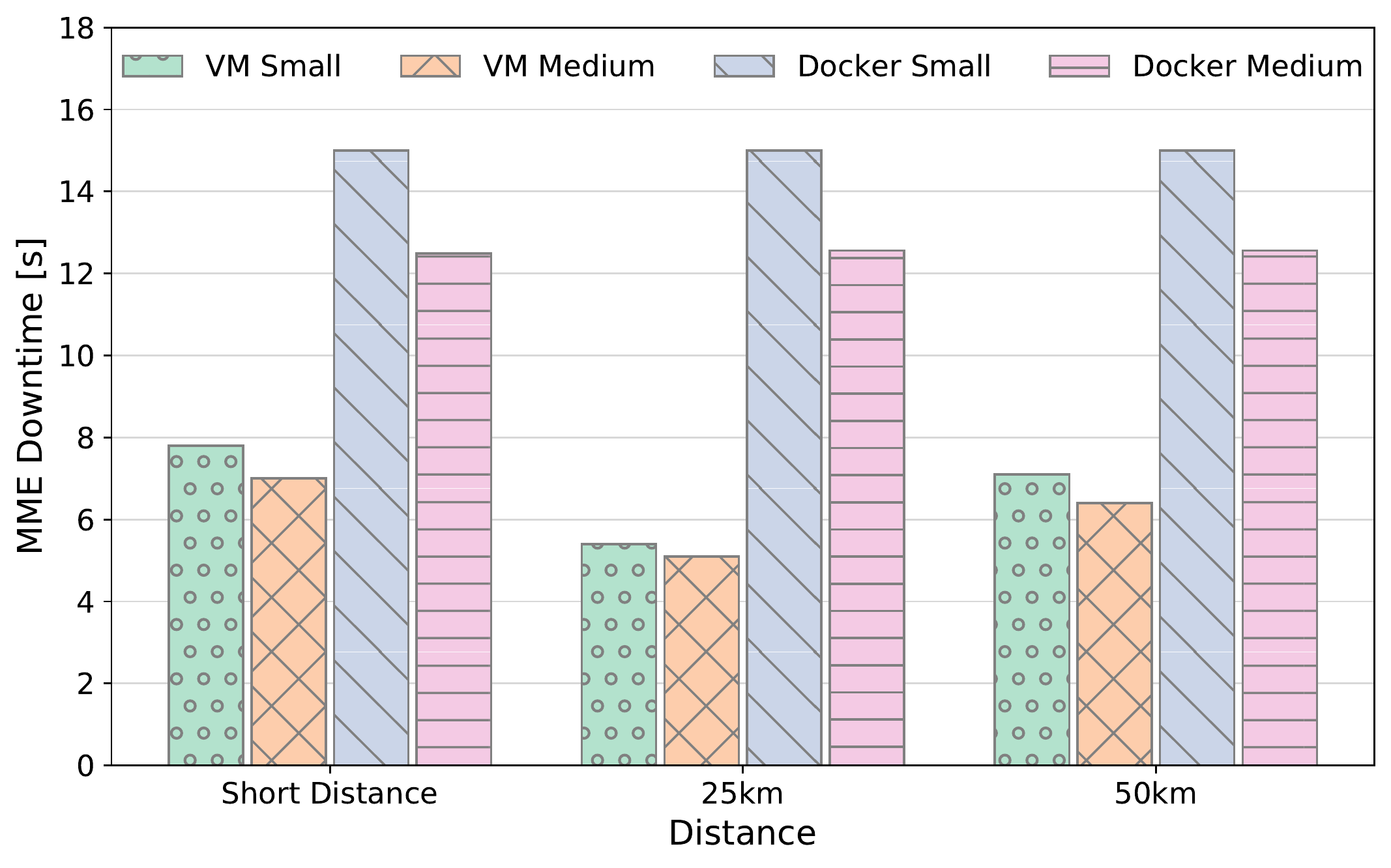}
    \caption{ SPGW - VM and Container downtime for three lightpath lengths and two flavor types.}
    \label{fig:SPGW-DownTime}
    \vspace{-1.2em}
    \vspace{1em}
\end{figure}

Fig.~\ref{fig:SPGW-DownTime} shows the downtime values during the SPGW migration in the OpenROADM testbed. 
More interestingly ---- and in contrast to the HSS and MME downtime analysis --- a modest downtime 
variation can be observed for the SPGW VM migration as the flavor size increases. 
This is mainly due to the combined influence of the dirty page contribution time and the virtual network interface reconfiguration. For the SPGW application, the main memory usage is intensified with the continuous uplink data request from the UE ping request. As the CPU cores increase from the flavor size upgrade, the dirty page copy time reduces, that eventually helped to reduce the SPGW downtime for the VM flavor size increase.

\begin{figure}[htbp]
     \vspace{-1.1em}
     \centering
    \includegraphics[trim=0 0 0 -1cm,width=1.0\columnwidth]{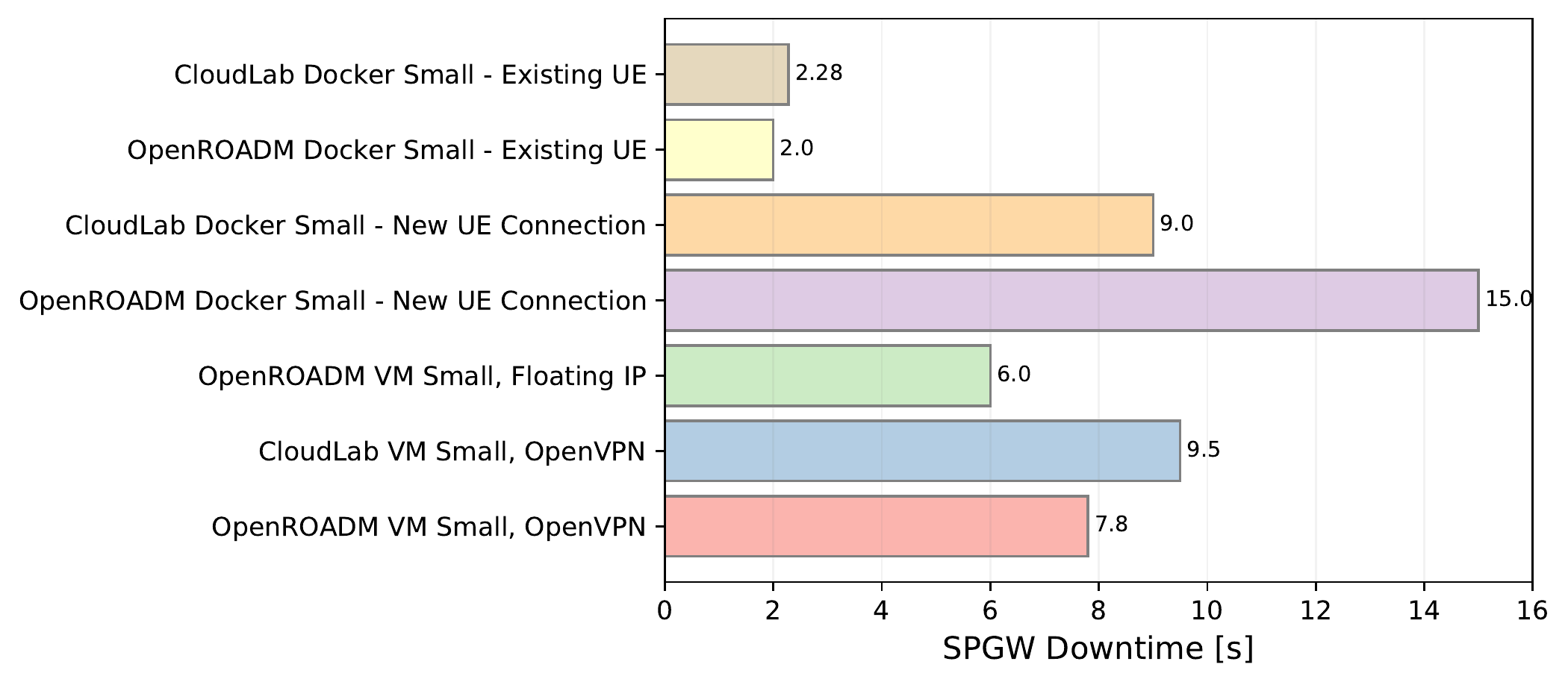}
    \caption{Comparison of SPGW downtime in the OpenROADM and CloudLab testbeds.}
    \label{fig:SPGW-CloudDownTime}
    \vspace{-1.8em}
    \vspace{1em}
\end{figure}

Fig.~\ref{fig:SPGW-CloudDownTime} compares the SPGW downtime collected in both the OpenROADM and CloudLab testbed. 
During the SPGW migration, the UE service is temporarily paused. In this case, the SPGW Container downtime is presented with additional cases: i) downtime for the existing UE (who has the user data traffic); ii) and downtime for the new UE connection (who initiates the attached procedure during the SPGW service unavailability). 
The downtime for the existing UE is significantly less than that of the new UE connection. This is because, the existing UE connectivity is re-established (with the newly contributed GTP utility program support specified in Sec.~\ref{sec:dc}) once the GTP tunneling information along with the network is restored at the destination node. 
However, in case of the new UE connection, the SPGW needs to be restored successfully with the control plane sockets update, which influences the increase in the downtime value for the Container.

\begin{figure}[htbp]
     \vspace{-1.1em}
     \centering
     \includegraphics[trim=0 0 0 -1cm,width=1.0\columnwidth]{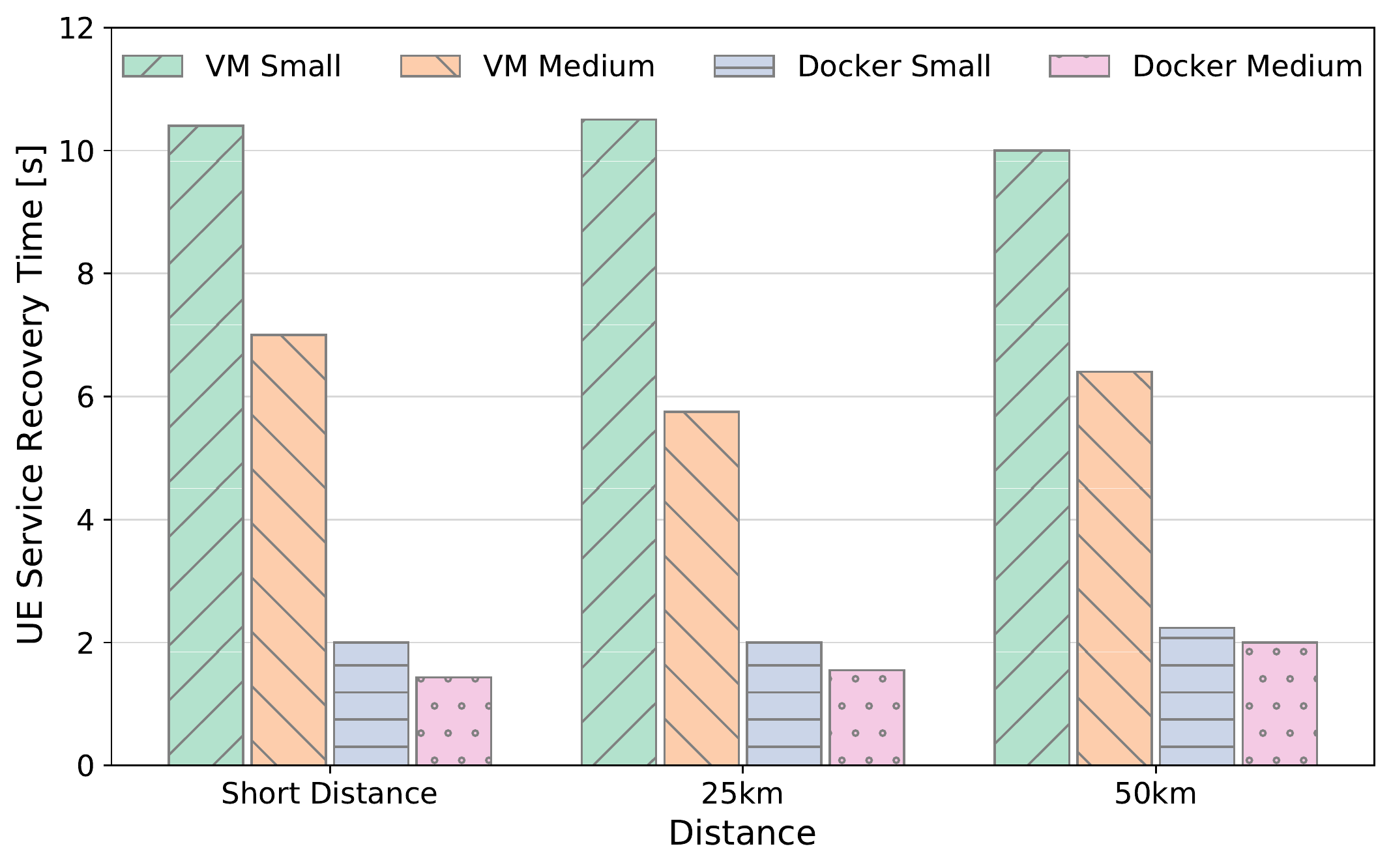}
    \caption{UE SRT for three lightpath lengths and two flavor types.}
    \label{fig:SPGW-UESRT}
    \vspace{-1.2em}
    \vspace{1em}
\end{figure}
\begin{figure}[htbp]
     \vspace{-1.1em}
     \centering
     \includegraphics[trim=0 0 0 -1cm,width=1.0\columnwidth]{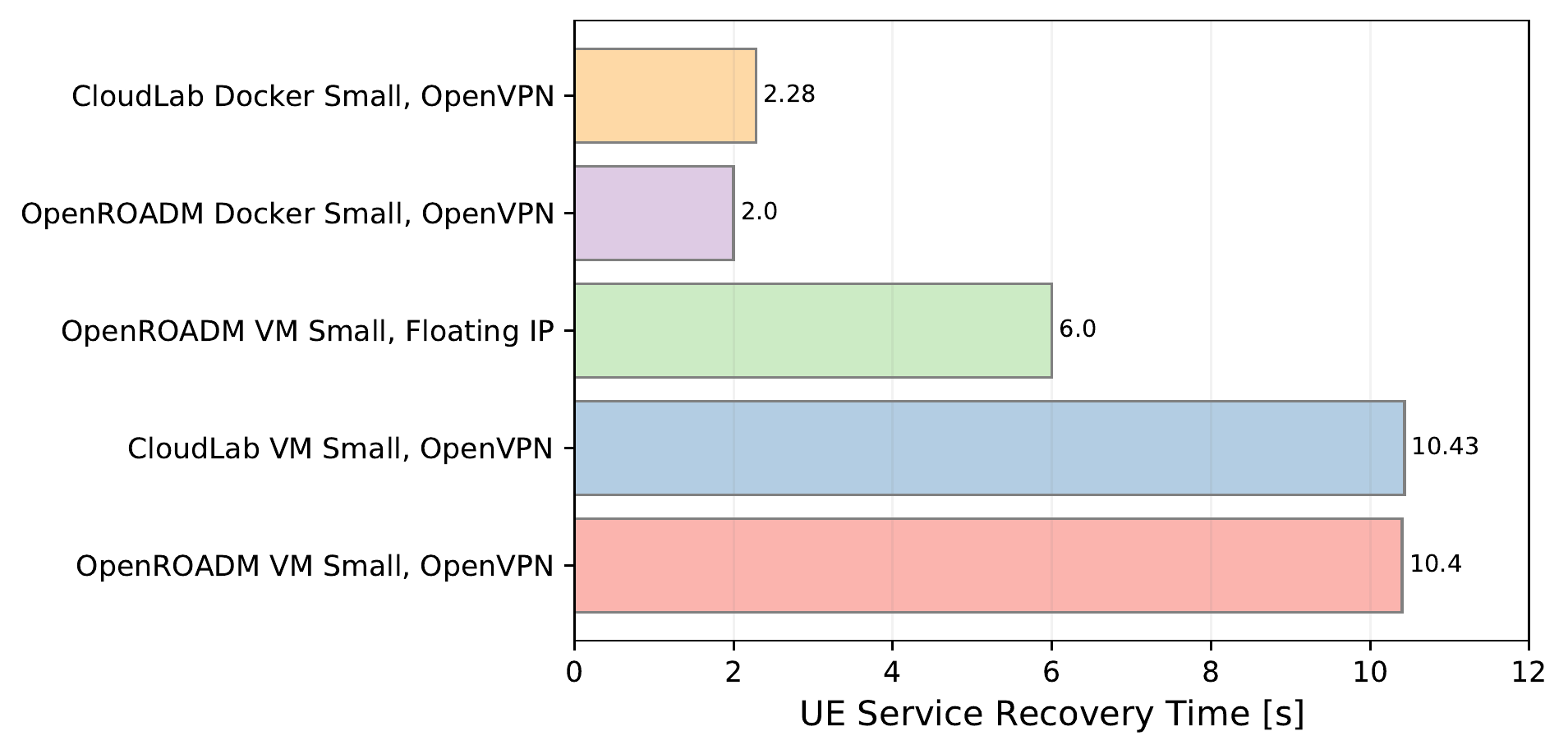}
    \caption{Comparison of UE SRT in OpenROADM and CloudLab testbeds.}
    \label{fig:SPGW-CloudUESRT}
    \vspace{-1.8em}
    \vspace{1em}
\end{figure}

Fig.~\ref{fig:SPGW-UESRT} reports the UE Service Recovery Time (SRT) of both the VM and Container during the SPGW migration in the OpenROADM testbed. 
The UE SRT value is less for the Container than that of the VM for all lightpath lengths and flavor types due to the newly contributed GTP utility software program at the Container. 
In addition, as the flavor size increases, the application performance improves to regain the UE connectivity faster for both the VMs and the Containers. 
Moreover, for the Container, no significant impact is observed in UE SRT due to the lightpath length change. This is because, the round trip time between the physical server and the VNFs in the compute node is less than one millisecond.

Fig.~\ref{fig:SPGW-CloudUESRT} compares the UE SRT measured in the two testbeds.
The SRT is 10 seconds when OpenVPN is used in both testbeds. 
The reason is that the network connectivity restoration time with 
OpenVPN takes about 8 seconds as depicted in Fig.~\ref{fig:SPGW-CloudDownTime}. 
After restoration, about 2 extra seconds are required to route the UDP encapsulated GTP traffic through the OpenVPN server.
The SRT reduces to 6 seconds when Floating IP is used in the OpenROADM testbed. 

\section{Conclusions and Future Study}

This paper reports the first set of public experiments about a NFV enabled mobile network comprising 
a backhaul fiber optics transport network that is entirely built with the latest OpenROADM 
compliant equipment and SDN control technology.
Through the single point of coordination provided by the PROnet Orchestrator module
--- for joint control of the backhaul optical layer, the Ethernet layer, and the compute resources ---
live migration of three EPC components --- HSS, MME, and SPGW --- 
virtualized through either VM or Container technology is experimentally 
achieved without permanent loss of UE connectivity.
To successfully carry out the Container live migration of the three EPC components, the authors designed and developed a number of custom functions 
that permit to overcome the limitations of both OAI and CRIU current open software packages.
These functions support migration of the end-points for both the GTP and SCTP 
connections that are employed by the SPGW and CU-MME, respectively.
The newly added software packages and upgrades are also tested on the federated CloudLab testbed,
which provides a third party and open platform for independent compliance validation of the said software.

Experimental results obtained using both testbeds (the OpenROADM and the CloudLab) are reported and discussed.
Specifically, migration time and service downtime performance indicators for
the two virtualization technologies (VM and Container) are compared,
while accounting for a number of system factors like flavor type of the
computing instances (compute, memory, and storage capacity), 
length of the temporary lightpath created between two compute sites and used to expedite the 
migration of the EPC component of interest, and type of network interface (OpenVPN and Floating IP) applied.
It is shown that fine tuning of these factors may be required to achieve optimal performance.

Outside the scope of this paper and possible subject of future studies,
live migration of virtualized CU/DU (vCU/vDU) using both the VM and Container technologies
is an additional critical functionality required in C-RAN.
With the 3GPP recommended functionality split options, the (OAI) vCU/vDU 
modules must cope with the backhaul transport network dependency and
meet the desired mobile network service latency and throughput. 
For example, CU interacts with MME using the SCTP transport layer protocol,
for which support can be provided in the CRIU code as described in this paper.
However, the CRIU code and related kernel level changes must be handled carefully 
to specifically account for the network  requirements (e.g., fronthaul latency) 
as dictated by the split option chosen for the vCU/vDU pair.
Another aspect that remains to be investigated is the possible reduction 
of Container service downtime by means of lazy migration technique. 

While a number of open challenges remains to be addressed before achieving a completely flexible virtualized C-RAN solution 
that is capable of supporting live migration of all of its components, the contribution of this paper takes
C-RAN a step closer to that ultimate goal of enabling power management, 
load balancing, and fault tolerance in the cloud environment assigned to support the needed NFV.

\section*{Acknowledgment}
This work is supported in part by NSF grants CNS-1405405, CNS-1409849, ACI-1541461, CNS-1531039T and also partially funded by the EU Commission through the 5GROWTH project (grant agreement no. 856709).

\bibliographystyle{IEEEtran}
\begin{sloppypar}
\bibliography{biblio}
\end{sloppypar}

\end{document}